# Combined DFT, SCAPS-1D, and wxAMPS frameworks for design optimization of efficient $Cs_2BiAgI_6$-based perovskite solar cells with different charge transport layers

M. Khalid Hossain[1,*], A.A. Arnab[2], Ranjit C. Das[3], K. M. Hossain[4], M. H. K. Rubel[4], Md. Ferdous Rahman[5], H. Bencherif[6], M. E. Emetere[7], Mustafa K. A. Mohammed[8], Rahul Pandey[9]

[1]*Institute of Electronics, AERE, Bangladesh Atomic Energy Commission, Dhaka 1349, Bangladesh*
[2]*Dept. of EEE, Ahsanullah University of Science and Technology, Dhaka 1208, Bangladesh*
[3]*Materials Science and Engineering, Florida State University, Tallahassee, FL 32306, USA*
[4]*Department of Materials Science and Engineering, University of Rajshahi, Rajshahi 6205, Bangladesh*
[5]*Department of Electrical and Electronic Engineering, Begum Rokeya University, Rangpur 5400, Bangladesh*
[6]*Higher National School of Renewable Energies, Environment and Sustainable Development, Batna 05078, Algeria*
[7]*Department of Physics and Solar Energy, Bowen University, Iwo 232101, Osun, Nigeria*
[8]*Radiological Techniques Department, Al-Mustaqbal University College, 51001 Hillah, Babylon, Iraq*
[9]*VLSI Centre of Excellence, Chitkara University, Punjab 140401, India*

Correspondence: *khalid.baec@gmail.com; khalid@kyudai.jp; ORCID: https://orcid.org/0000-0003-4595-6367

**Abstract**

In this study, combined DFT, SCAPS-1D, and wxAMPS frameworks are used to investigate the optimized designs of $Cs_2BiAgI_6$ double perovskite-based solar cells. The first-principle calculation is employed to investigate the structural stability, optical responses, and electronic contribution of the constituent elements in $Cs_2BiAgI_6$ absorber material, where SCAPS-1D and wxAMPS simulators are used to scrutinize different configurations of $Cs_2BiAgI_6$ solar cells. Here, PCBM, ZnO, $TiO_2$, $C_{60}$, IGZO, $SnO_2$, $WS_2$, and $CeO_2$ are used as ETL, and $Cu_2O$, CuSCN, $CuSbS_2$, NiO, P3HT, PEDOT: PSS, Spiro-MeOTAD, CuI, CuO, $V_2O_5$, CBTS, CFTS are used as HTL, and Au is used as a back contact. About ninety-six combinations of $Cs_2BiAgI_6$-based solar cell structures are investigated, in which eight sets of solar cell structures are identified as the most efficient structures. Besides, holistic investigation on the effect of different factors such as the thickness of different layers, series and shunt resistances, temperature, capacitance, Mott-Schottky and generation-recombination rates, and *J-V* (current-voltage density) and *QE* (quantum efficiency) characteristics is performed. The results show CBTS as the best HTL for $Cs_2BiAgI_6$ with all eight ETLs used in this work, resulting in a power conversion efficiency (PCE) of 19.99%, 21.55%, 21.59%, 17.47%, 20.42%, 21.52%, 14.44%, 21.43% with PCBM, $TiO_2$, ZnO, $C_{60}$, IGZO, $SnO_2$, $CeO_2$, $WS_2$, respectively. The proposed strategy may pave the way for further design optimization of lead-free double perovskite solar cells.

## 1 Introduction

The depletion of energy sources and consumption of fossil fuels, natural gas, coal, etc., have a negative impact on our society and environment [1–5]. Therefore, developing alternative energy resources and sustainable energy sources with economic and environmental concerns are the center of research worldwide [6–8]. A solar cell is one of the most potential renewable and sustainable energy sources to attain the increasing energy demand and mitigate global warming [6]. In recent years, Perovskite solar cells (PSCs) have attracted significant research attention as a new approach in solar photovoltaic (PV) technology due to their significant power conversion efficiency (PCE) improvement from 3.8% to 25.2% since 2009 [9,10]. The general formula of PSCs is $ABX_3$, where A and B represent monovalent cation (Methylammonium or $CH_3NH_3^+$, Formamidinium or $H_2NCHNH_2^+$, $Cs^+$, $CHCH_3^+$) and divalent metal cation (Pb, Sn, Ti, Bi, Ag), respectively, while X represents halogen anion ($Cl^-$, $Br^-$ and $I^-$). Compared to the other PSCs, lead-halide and organic-inorganic hybrid PSCs have received enormous attention due to their higher power conversion efficiency (PCE) of around 25%, which is comparable to the champion silicon solar cell that has a record PCE of 26.7% [11–13]. Despite the outstanding performance of PSCs, two major challenges prevent them from commercial applications like the use of toxic material lead and organic cations, which have volatile and hygroscopic nature [14]. Organic-inorganic hybrid PSCs are chemically unstable when exposed to oxygen, moisture, and high temperatures because the organic cations are hygroscopic and volatile [15]. On the other hand, lead-based halide perovskite is a highly toxic and hazardous material that can cause severe health and environmental issues [16–20].

As a result, there is a growing demand for nontoxic lead-free PSCs, leading researchers to develop several lead-free perovskite materials. As an alternative to lead-based perovskites, $Pb^{2+}$ has been replaced by various nontoxic elements, including bivalent $Sn^{2+}$ and $Ge^{2+}$ [21]. However, $Sn^{2+}$ and $Ge^{2+}$ in lead-free PSCs demonstrate low stability due to oxidation [21]. Also, $Pb^{2+}$ has been replaced with heterovalent $M^{3+,}$ such as $Bi^{3+}$, which is non-toxic, isoelectronic with $Pb^{2+}$, and stable semiconducting halides [22]. Bismuth-based PSCs exhibit longer charge carrier diffusion lengths due to their lower intrinsic trap densities and defect states [23]. However, when a highly charged $Bi^{3+}$ ion was added to the three-dimensional $A^{1+}M^{2+}X_3$ structure, the three-dimensional $A^{1+}M^{2+}X_3$ structure produced poor optoelectronic characteristics compared to the lead-based perovskite [24,25]. To overcome these undesirable properties, the Elpasolite structure, also known as the double perovskite structure, has been used by adding $Bi^{3+}$ anion [26]. The general formula of Elpasolite structure is $A_2M^{1+}M^{3+}X_6$, where A, X, $M^{1+}$, and $M^{3+}$ stand for monovalent cation, halide anion ($Br^-$, $Cl^-$, $I^-$), inorganic cation ($Cu^+$, $Ag^+$, $Au^+$, $Na^+$, $K^+$, $Rb^+$, and $In^+$), and organic or inorganic cation ($Bi^{3+}$ or $Sb^{3+}$), respectively [26]. Recent studies showed that $Bi^{3+}$-based double perovskites with monovalent cation $Ag^{1+}$ are a very promising material for photovoltaic applications due to their desirable band gap, comparable charge carrier effective masses, excellent photoluminescence lifetime, extended carrier recombination lifetimes, and high stability [27–33]. McClure and his coworker reported that $Cs_2AgBiBr_6$ and $Cs_2AgBiCl_6$ have an excellent band gap and high stability compared to the $CH_3NH_3PbX_3$. However, $Cs_2AgBiBr_6$ and $Cs_2AgBiCl_6$ are exhibited low efficiency, as large charge carrier effective masses, low charge carrier transport capabilities, and high band gap (>2 eV) [34–36], which make them unfit for solar cells. On the contrary, the $Cs_2AgBiI_6$ absorber exhibited a favorable band gap (1.12 eV), higher light absorbing capacities, and higher performance than $Cs_2AgBiBr_6$ and $Cs_2AgBiCl_6$ which makes it suitable for the PSC [37,38].

The performance and efficiency of the absorber can be optimized in PSCs by using a suitable electron transport layer (ETL) and a hole transport layer (HTL). The properties and type of layer material used in PSCs significantly impact stability and performance. When choosing an HTL for the PSCs, the material properties should be considered, including the valence band offset between absorber and HTL, hole mobility, and cost [39]. Researchers mostly used HTLs such as Spiro-MeOTAD and PEDOT: PSS in PSCs due to their excellent tunability and processability [40,41]. However, the challenges associated with these materials are the processing cost, poor conductivity and hole transport, and stability [42]. On the other hand, the ETL should have a conduction band offset between the absorber and ETL, compatible with other layers' high electron mobility, and cost efficiency [39]. The maximum PCE of 28.4% is exhibited when $TiO_2$ as ETL is used in PSCs with $MASnI_3$ as the absorber layer [43]. However, the rutile crystalline phase of $TiO_2$, which needs to be processed at high annealing temperatures, is used for solar cell applications [44–51]. In addition, mesoporous-$TiO_2$ as ETL causes PSC's environment stability to deteriorate when exposed to ultraviolet light [44]. However, the low atmospheric stability and high-temperature processing of $TiO_2$ constrained them as ETL from PSCs. So far, a few studies have been conducted to enhance the device performance by selecting HTL and ETL with absorber layers in PSCs [52,53]. In our recent study, we reported that $CsPbI_3$ absorber layer-based device with CBTS as HTL exhibited PCE of 16.71%, 17.90%, 17.86%, 14.47%, 17.76%, 17.82% with PCBM, $TiO_2$, ZnO, $C_{60}$, IGZO, $WS_2$ ETL respectively [14]. Another study demonstrates that using $SnO_2$ as ETL and CuSCN as HTL with $MAPbI_3$ absorber demonstrated a high PCE of 27% [52]. Therefore, more research needs to be done to investigate the device performance of PSCs for different ETL, HTL, and back contact metals with lead-free perovskite absorber layers, to outperform Shockley Queisser's (SQ) efficiency limit for a single solar cell.

The electronic properties such as band gap and band structure, density of states (DOS), and charge density distribution are crucial for the investigation of the electronic contribution of different elements in perovskites and other structural compounds [14,54,55]. The state of a material system can be properly understood by exploring its physical properties. It can also identify a material's possible practical applications. Nowadays, researchers have been performing theoretical works to explore the physical properties of materials of interest via the density functional theory (DFT). According to reports [14,56–58], several halide perovskite materials exhibit intriguing physical characteristics including structural, electrical, optical, and mechanical characteristics that make them potential candidates for optoelectronic and photovoltaic operations. Recently, Hadi *et al.* [59] have investigated the structural, electronic, optical, and mechanical properties of lead-free cubic double perovskite-type $Cs_2AgBiBr_6$ using the DFT method. The indirect band gap of $Cs_2AgBiBr_6$ was reduced and switched to direct by making the compound disordered following the disordering of $Ag^+/Bi^{3+}$ cations by creating an antisite defect in the sublattice. This reduction of the band gap is responsible to enhance the optical absorption in the visible region and makes $Cs_2AgBiBr_6$ suitable for solar cell applications [59]. Besides, the electronic band structure of isostructural double perovskite halide $Cs_2AgSbCl_6$ has also been investigated by both the PBE and HSE approximations [60]. It was reported that $Cs_2AgSbCl_6$ has an indirect band gap of 1.40 eV (HSE) / 2.35 eV (PBE). To the best of our knowledge, the theoretical research on various physical properties of our chosen double perovskite-type halide $Cs_2AgBiI_6$ is not reported yet. Therefore, it is necessary to discover the physical properties of $Cs_2AgBiI_6$ system using computational studies like DFT.

In the present study, the performance of the lead-free $Cs_2AgBiI_6$ halide PSC is investigated through the one-dimensional solar cell capacitance simulator (SCAPS-1D) platform for the very first time using numerous ETLs and HTLs. During the study, the performance is evaluated through $TiO_2$, PCBM, ZnO, $C_{60}$, IGZO, $SnO_2$, $WS_2$, $CeO_2$ as ETL, and $Cu_2O$, CuSCN, $CuSbS_2$, NiO, P3HT, PEDOT: PSS, Spiro-MeOTAD, CuI, CuO, $V_2O_5$, CBTS, CFTS as HTL with Gold (Au) as back contact metal (**Figure 1(a)**). Furthermore, the band gap of the $Cs_2BiAgI_6$ double perovskite absorber is also validated using theoretical first-principle calculations via the DFT framework together with the investigation of its structural and optical properties for the first time. In addition, we investigated the performance of the ETL and HTL layer along with the effect of absorber and ETL thickness, series resistance, shunt resistance, working temperature, capacitance and Mott-Schottky, generation and recombination rate, and current-voltage density (*J-V*) and quantum efficiency (*QE*) characteristics. Finally, the performance of the best configurations was verified through wxAMPS (widget-provided analysis of microelectronic and photonic structures) simulation findings.

## 2 Materials and methodology

### 2.1 First principal calculations of $Cs_2BiAgI_6$ absorber using DFT

This study performs the first-principle calculations exploiting the Cambridge Serial Total Energy Package (CASTEP) [61] engaged in density functional theory (DFT) [62], incorporating the principle of pseudo-potential plane-wave (PP-PW) total energy calculation to identify the ground state structure of $Cs_2AgBiI_6$. To grow electron-ion interactions, the Vanderbilt-type ultrasoft pseudopotential is selected [63]. The exchange-correlation potential is enacted through the Generalized Gradient Approximation (GGA) composed with the Perdew-Burke-Ernzerhof (PBE) functional [64]. For ensuring the convergence the cut-off energy of 520 eV is applied. The Monkhorst-Pack scheme [65] is employed by taking a *k*-point mesh of 7 × 7 × 7, which is settled to an ultrafine mode for the first Brillouin zone. However, to see clearly the electronic charge density pattern, a larger size *k*-point grid (12 × 12 × 12) was used. It is also essential to apply the algorithm of Broyden-Fletcher-Goldfarb-Shanno (BFGS) during structural optimization [66] so that the relatively exact lattice constants and internal atomic coordinates of $Cs_2AgBiI_6$ are assessed. It is usual to be relaxed the lattice parameters and atomic positions after structural optimization [67]. The optimum converging functions to obtain the ground state structure are selected as a maximum force within 0.03 eV, maximum ionic displacement within 0.001Å, total energy difference within $1 \times 10^{-5}$ eV per atom, and maximum stress within 0.05 GPa. The electronic and optical properties are then calculated for the optimized structure by considering the aforementioned parameters.

### 2.2 SCAPS-1D numerical simulation

SCAPS-1D carried out the device modeling and simulation. The continuity, drift-diffusion, and position-dependent Poisson equations provide the backbone of the simulation. The three primary semiconductor equations used in the simulation framework are the continuity equations for holes (**Eq. 1**), electrons (**Eq. 2**), and the Poisson equation (**Eq. 3**), as shown below [68,69].

$$\frac{dP_n}{dt} = G_{P} - \frac{P_n - P_{n0}}{\tau_p} - P_n u_P \frac{dE}{dx} - u_P E \frac{dP_n}{dX} + D_P \frac{d^2 P_n}{dx^2} \tag{1}$$

$$\frac{dn_P}{dt} = G_n - \frac{n_P - n_{P0}}{\tau_n} - n_P u_n \frac{dE}{dx} - u_n E \frac{dn_P}{dx} + D_P \frac{d^2 n_P}{dx^2} \tag{2}$$

$$\frac{d}{dX}\left(\varepsilon(x)\frac{d\emptyset}{dx}\right) = q[\, p(x) - n(x) + N_{d^+}(x) - N_{a^-}(x) + p_t(x) - n_t(x)] \quad (3)$$

Where q stands for electron charge, $\varepsilon$ stands for dielectric permittivity, G stands for the rate of generation, D stands for the diffusion coefficient, $\emptyset$ stands for electrostatic potential, E stands for the electric field, p(x) stands for the number of free holes, n(x) stands for the number of free electrons, $p_t$ (x) stands for the number of trapped holes, $n_t$(x) stands for the number of trapped electrons, $N_{d+}$ stands for the concentration of donor-ionized doping, and $N_a-$ stands for the concentration of acceptor ionized doping, and x stands for the thickness.

## 2.3 wxAMPS numerical simulation

The wxAMPS, a program created by the University of Illinois, was used to do a similar numerical analysis as a secondary verification of solar cells' performance in this work [70,71]. The device behavior is represented via resolving Poisson's equation (**Eq. 4**), which links the electrostatic potential to charge, as well as the electron and hole continuity equations (**Eqs 5** and **6**) in the different regions of the device structure.

$$\frac{\delta}{\delta x}\left(\varepsilon(x)\frac{\delta(\phi(x))}{\delta x}\right) = e\rho \quad (4)$$

$$\frac{1}{e}\frac{\delta J_n}{\delta x} = G(x) - R(x) \quad (5)$$

$$\frac{1}{e}\frac{\delta J_P}{\delta x} = -G(x) + R(x) \quad (6)$$

Where $x$ denotes the position, $\varepsilon$ denotes the dielectric constant, $\phi$ denotes the local electric potential, e denotes the electron charge, $\rho$ denotes the summed charge density, $J_n$ denotes the electron current density, $J_p$ denotes the hole electron density, G denotes the optical carrier generation rate, and R denotes the overall charge carrier recombination rate. The AMPS-1D program employs two distinct models: the lifetime model and the density of state (DOS) model. For a clear comprehension, the recombination phenomena, and changes in defect states, as well as their effects on the electric field variance across the charges transport materials/perovskite interfaces, were taken into consideration in this study using the density of state model. By resolving three-coupled non-linear differential equations with appropriate boundary conditions, the AMPS-1D can determine the quasi-Fermi level and the electrostatic potential at all locations in the device. The solar cell figures of merit can be computed in accordance with the definitions of these variables based on the device depth.

## 2.4 $Cs_2BiAgI_6$-based PSC structure

Here, along with the $Cs_2BiAgI_6$ absorber layer, ETL, HTL, and back contact are associated to form the structure of double perovskite solar cell. The solar cell structure of the $Cs_2BiAgI_6$ absorber forms an n-i-p structure. In this instance, an n-i-p structure's superior long-wavelength response makes it preferable to a traditional semiconductor p-n junction, and deep within the device, across the intrinsic area, the depletion zone of an n-i-p structure is present. When a cell is exposed to long wavelength radiation, photons enter the cell deeply. However, only electron-hole pairs produced in and close to the depletion area can contribute to the creation of current. An efficient creation and separation mechanisms of electron-hole pairs are made possible by the greater depletion width, which raises the cell's quantum efficiency [14]. $Cs_2BiAgI_6$ captures photons due to double heterostructure it ensures the charge and photon confinement and both sides of highly doped ETL and HTL work as an ohmic contact. During the study SCAPS-1D, the software helps us to investigate the performance of various structures of double PSCs. By keeping the ambient temperature 300 K, frequency 1 MHz, and AM 1.5 G sunlight spectrum double perovskite structures are formed. Also taken eight ETLs and twelve HTLs, back contact as Au to investigate the different structures and their optoelectronic parameters are set up initially from the different studies which are mentioned in **Tables 1- 3**.

**Table 1**. Input optimization parameters of the TCO, ETL, and absorber layer of the study.

| **Parameters** | **ITO** | **$TiO_2$** | **PCBM** | **ZnO** | **$C_{60}$** | **IGZO** | **$SnO_2$** | **$WS_2$** | **$CeO_2$** | **$Cs_2BiAgI_6$** |
|---|---|---|---|---|---|---|---|---|---|---|
| Thickness (nm) | 500 | 30 | 50 | 50 | 50 | 30 | 100 | 100 | 100 | 800* |
| Band gap, $E_g$ (eV) | 3.5 | 3.2 | 2 | 3.3 | 1.7 | 3.05 | 3.6 | 1.8 | 3.5 | 1.6 |
| Electron affinity, X (eV) | 4 | 4 | 3.9 | 4 | 3.9 | 4.16 | 4 | 3.95 | 4.6 | 3.90 |
| Dielectric permittivity (relative), $\varepsilon_r$ | 9 | 9 | 3.9 | 9 | 4.2 | 10 | 9 | 13.6 | 9 | 6.5 |
| CB effective density of states, $N_C$(1/cm$^3$) | $2.2 \times 10^{18}$ | $2 \times 10^{18}$ | $2.5 \times 10^{21}$ | $3.7\times10^{18}$ | $8.0 \times 10^{19}$ | $5 \times 10^{18}$ | $2.2 \times 10^{18}$ | $1 \times 10^{18}$ | $1 \times 10^{20}$ | $1\times 10^{19}$ |
| VB effective density of states, $N_V$ (1/cm$^3$) | $1.8 \times 10^{19}$ | $1.8 \times 10^{19}$ | $2.5 \times 10^{21}$ | $1.8\times10^{19}$ | $8.0 \times 10^{19}$ | $5 \times 10^{18}$ | $1.8 \times 10^{19}$ | $2.4 \times 10^{19}$ | $2 \times 10^{21}$ | $1 \times 10^{19}$ |
| Electron mobility, $\mu_n$ (cm$^2$/Vs) | 20 | 20 | 0.2 | 100 | $8.0 \times 10^{-2}$ | 15 | 100 | 100 | 100 | 2 |
| Hole mobility, $\mu_h$ (cm$^2$/Vs) | 10 | 10 | 0.2 | 25 | $3.5 \times 10^{-3}$ | 0.1 | 25 | 100 | 25 | 2 |
| Shallow uniform acceptor density, $N_A$ (1/cm$^3$) | 0 | 0 | 0 | 0 | 0 | 0 | 0 | 0 | 0 | $1 \times 10^{15*}$ |
| Shallow uniform donor density, $N_D$ (1/cm$^3$) | $1\times10^{21}$ | $9 \times 10^{16}$ | $2.93 \times 10^{17}$ | $1 \times 10^{18}$ | $1 \times 10^{17}$ | $1 \times 10^{17}$ | $1 \times 10^{17}$ | $1\times10^{18}$ | $10^{21}$ | 0* |
| Defect density, $N_t$ (1/cm$^3$) | $1 \times 10^{15*}$ | $1 \times 10^{15*}$ | $1 \times 10^{15*}$ | $1 \times 10^{15*}$ | $1 \times 10^{15*}$ | $1 \times 10^{15*}$ | $1 \times 10^{15*}$ | $1\times10^{15*}$ | $1 \times 10^{15*}$ | $1\times10^{15*}$ |
| Reference | [14] | [14] | [14] | [14] | [14] | [14] | [14] | [14] | [14] | [72] |

*This study

**Table 2**. Input optimization parameters HTL the study [14].

| HTL | $Cu_2O$ | CuSCN | $CuSbS_2$ | P3HT | PEDOT: PSS | Spiro-MeOTAD | NiO | CuI | CuO | $V_2O_5$ | CFTS | CBTS |
|---|---|---|---|---|---|---|---|---|---|---|---|---|
| Thickness (nm) | 50 | 50 | 50 | 50 | 50 | 200 | 100 | 100 | 50 | 100 | 100 | 100 |
| Band gap, Eg (eV) | 2.2 | 3.6 | 1.58 | 1.7 | 1.6 | 3 | 3.8 | 3.1 | 1.51 | 2.20 | 1.3 | 1.9 |
| Electron affinity, X (eV) | 3.4 | 1.7 | 4.2 | 3.5 | 3.4 | 2.2 | 1.46 | 2.1 | 4.07 | 4.00 | 3.3 | 3.6 |
| Dielectric permittivity (relative), εr | 7.5 | 10 | 14.6 | 3 | 3 | 3 | 10.7 | 6.5 | 18.1 | 10.00 | 9 | 5.4 |
| CB effective density of states, NC ($1/cm^3$) | $2 \times 10^{19}$ | $2.2 \times 10^{19}$ | $2 \times 10^{18}$ | $2 \times 10^{21}$ | $2.2 \times 10^{18}$ | $2.2 \times 10^{18}$ | $2.8 \times 10^{19}$ | $2.8 \times 10^{19}$ | $2.2 \times 10^{19}$ | $9.2 \times 10^{17}$ | $2.2 x 10^{18}$ | $2.2 x 10^{18}$ |
| VB effective density of states, NV ($1/cm^3$) | $1 \times 10^{19}$ | $1.8 \times 10^{18}$ | $1 \times 10^{1}$ | $2 \times 10^{21}$ | $1.8 \times 10^{19}$ | $1.8 \times 10^{19}$ | $1 \times 10^{19}$ | $1 \times 10^{19}$ | $5.5 \times 10^{20}$ | $5.0 \times 10^{18}$ | $1.8 x 10^{19}$ | $1.8 x 10^{19}$ |
| Electron mobility, µn (cm2/Vs) | 200 | 100 | 49 | $1.8 \times 10^{-3}$ | $4.5\times10^{-2}$ | $2.1\times10^{-3}$ | 12 | 100 | 100 | $3.2\times10^{2}$ | 21.98 | 30 |
| Hole mobility, µh ($cm^2/Vs$) | 8600 | 25 | 49 | $1.86 \times10^{-2}$ | $4.5\times10^{-2}$ | $2.16\times10^{-3}$ | 2.8 | 43.9 | 0.1 | $4.0\times10^{1}$ | 21.98 | 10 |
| Shallow uniform acceptor density, $N_A$ ($1/cm^3$) | $1 \times 10^{18}$ | $1 \times 10^{18}$ | $1 \times 10^{18}$ | $1 \times 10^{18}$ | $1 \times 10^{18}$ | $1.0 \times 10^{18}$ | $1 \times 10^{18}$ | $1.0 \times 10^{18}$ | $1 \times 10^{18}$ | $1 \times 10^{18}$ | $1 \times 10^{18}$ | $1 \times 10^{18}$ |
| Shallow uniform donor density, $N_D$ ($1/cm^3$) | 0 | 0 | 0 | 0 | 0 | 0 | 0 | 0 | 0 | 0 | 0 | 0 |
| Defect density, $N_t$ ($1/cm^3$) | $1.0\times10^{15*}$ | $1 \times 10^{15*}$ | $1 \times 10^{15*}$ | $1 \times 10^{15\ *}$ | $1 \times 10^{15*}$ | $1.0 \times 10^{15*}$ | $1 \times 10^{15*}$ | $1.0 \times 10^{15*}$ | $1 \times 10^{15*}$ | $1 \times 10^{15*}$ | $1 \times 10^{15*}$ | $1 \times 10^{15*}$ |

*This study

**Table 3**. Input parameters of interface defect layers [14].

| **Interface** | **Defect type** | **Capture Cross Section: Electrons/holes ($cm^2$)** | **Energetic Distribution** | **Reference for defect energy level** | **Total density ($cm^{-3}$) (integrated over all energies)** |
|---|---|---|---|---|---|
| ETL/$Cs_2BiAgI_6$ | Neutral | $1.0 \times 10^{-17}$<br>$1.0 \times 10^{-18}$ | Single | Above the VB maximum | $1.0 \times 10^{10}$ |
| $Cs_2BiAgI_6$ /HTL | Neutral | $1.0 \times 10^{-18}$<br>$1.0 \times 10^{-19}$ | Single | Above the VB maximum | $1.0 \times 10^{10}$ |

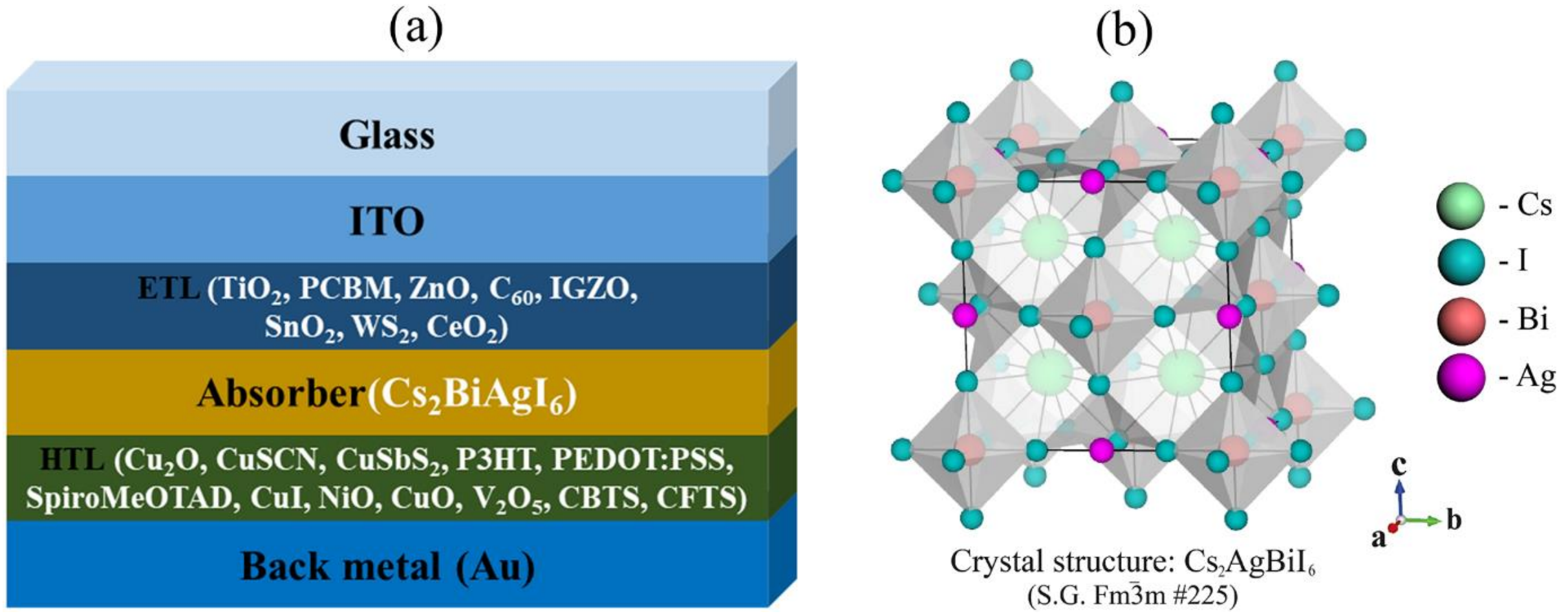

**Figure 1**. (a) The design configuration of the $Cs_2AgBiI_6$-based PSC, and (b) the crystal structure of the $Cs_2AgBiI_6$ cubic-single-perovskite semiconductor.

## 3 Results and discussion

### 3.1 Analysis of DFT results

#### *3.1.1 Structural properties of $Cs_2BiAgI_6$ compound*

According to the optimization output, the crystal structure of $Cs_2AgBiI_6$ is a cubic double-perovskite-type system with space group $Fm\bar{3}m$ (#225). The atomic positions of Cs, Ag, Bi, and I in the unit cell are 8c (0.25, 0.25, 0.25), 4a (0, 0, 0), 4b (0. 5, 0.5, 0.5), and 24e (0.2513, 0, 0). The optimized three-dimensional crystal structure of $Cs_2AgBiI_6$ is projected in **Figure 1(b)**. The calculated lattice parameter $a$ = 8.6338 Å is lower than that of the previously reported lattice parameter of similar double perovskite-type compound $Cs_2AgBiBr_6$ ($a$ = 11.430 Å) [59] due to the difference of atomic radius between I and Br as well as using different calculation parameters. However, the negative value of formation energy ($\Delta E_f$ = – 2.66 eV/atom) calculated by the following equation (**Eq. 7**) confirms the thermodynamic stability of optimized $Cs_2AgBiI_6$ [73].

$$\Delta E_f\,(\mathrm{Cs_2AgBiI_6}) = \frac{[E_{tot.}(\mathrm{Cs_2AgBiI_6}) - 2E_s(\mathrm{Cs}) - E_s(\mathrm{Ag}) - E_s(\mathrm{Bi}) - 6E_s(\mathrm{I})]}{N} \quad (7)$$

Here, $E_s$ (Cs), $E_s$(Ag), $E_s$(Bi), and $E_s$(I) are the energy of Cs, Ag, Bi, and I atoms, respectively, while $E_{tot}(Cs_2AgBiI_6)$ is the unit cell total energy of $Cs_2AgBiI_6$, and $N$ represents the number of atoms in the unit cell.

#### *3.1.2 Band structure and DOS of $Cs_2BiAgI_6$ compound*

The electronic properties provide important information to explain materials' bonding nature, photon-absorbing behaviours, and other relevant properties [74]. A compound's electronic properties are mostly related to its band structure, the density of states (DOS), and charge density. The result of band structure calculation along the highly symmetric directions within the $k$-space for $Cs_2AgBiI_6$ is presented in **Figure 2(a)**, where a red horizontal dash line at 0 eV indicates the Fermi level ($E_F$). The bands below and above the $E_F$ are usually called the valence and conduction bands, respectively. It is observed from **Figure 2(a)** that either the valence or conduction band is not crossed the $E_F$ and overlaps with each other. Hence the compound $Cs_2AgBiI_6$ has a band gap in its electronic band structure. The calculated band gap is found to be 0.846 eV. The band gap value of the studied compound is somewhat lower than that of other similar double perovskite-type compounds $Cs_2AgBiBr_6$ (2.19 eV) [26], $Cs_2AgBiCl_6$ (2.77 eV) [26], $Cs_2AgInCl_6$ (3.23 eV) [75], and $Cs_2AgSbCl_6$ (1.40 eV) [60].

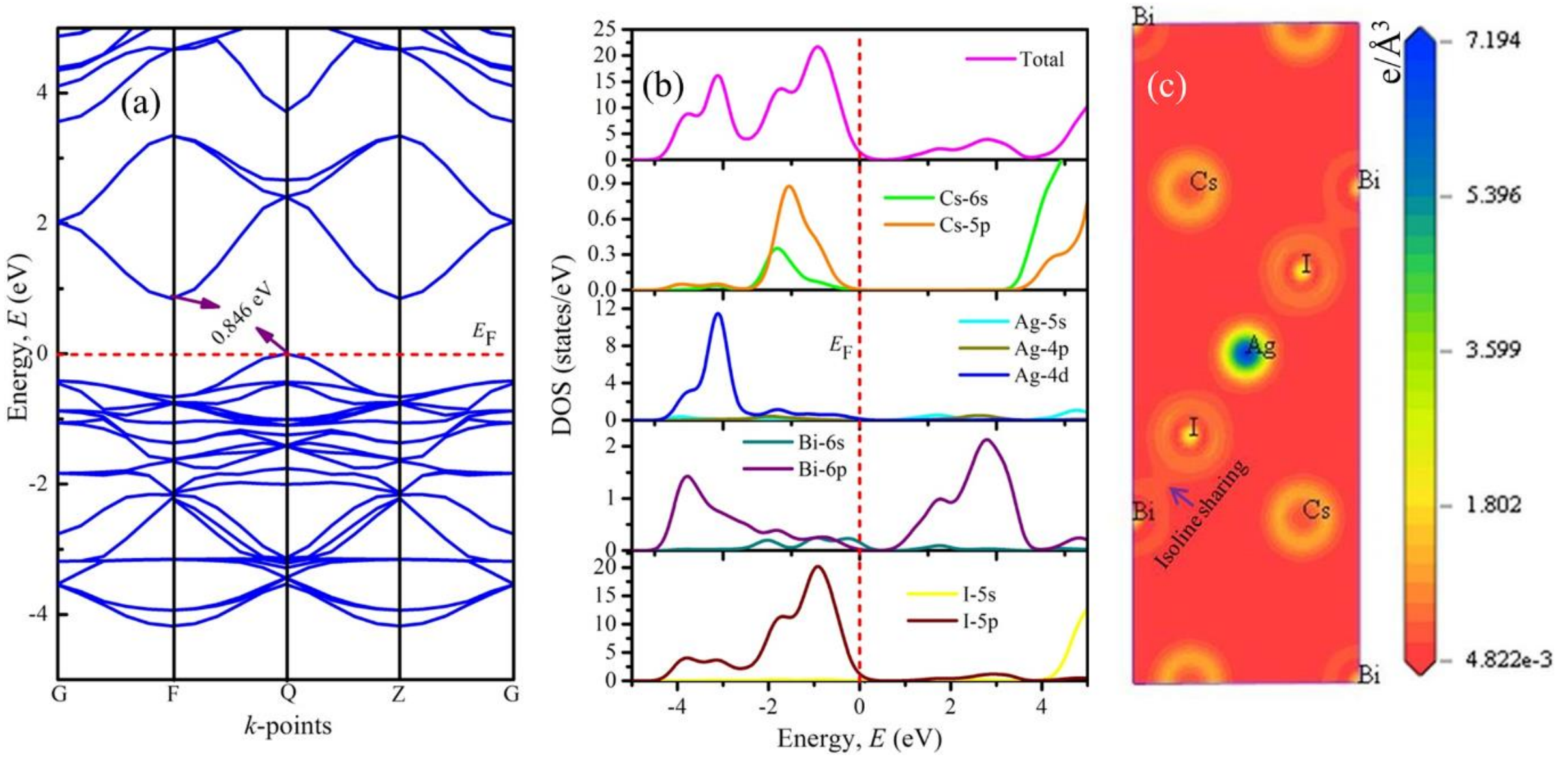

**Figure 2.** (a) Calculated band structure of $Cs_2AgBiI_6$ along the high-symmetry direction, (b) Calculated total and partial density of states of $Cs_2AgBiI_6$, and (c) The charge density mapping of $Cs_2AgBiI_6$ along (111) crystallographic plane.

Moreover, the minimum conduction band and maximum valence band are observed at F- and Q-points, respectively, exhibiting the indirect semiconducting nature of double perovskite compound $Cs_2AgBiI_6$. Besides, the valence band maximum is the nearest band to the $E_F$, which ensures the *p*-type semiconducting nature of chosen double perovskite. Interestingly, some nearly flat valence bands between F- and Z-points close to the $E_F$ are also observed. This highly dispersive nature of the bands near the $E_F$ indicated the high mobility of charge carriers and predicted that the effective mass would be significantly higher in these bands.

**Figure 2(b)** represents the total density of state (TDOS) together with the partial density of state (PDOS), which are used to understand further the contributions of different atoms/orbitals and the chemical bonding nature in $Cs_2AgBiI_6$. In the DOS diagram, the $E_F$ is represented by the red vertical dashed line fitted at 0 eV. The TDOS diagram confirms the band gap in the electronic structure of the studied double perovskite, reflecting a similar nature as the band diagram. The valence band mainly consists of Ag-4d and I-5p states together with the minimal contributions of Cs-6s, Cs-5p, and Bi-6p orbitals. On the other hand, the conduction band in the vicinity of the $E_F$ mostly arises from the Bi-6p state. However, the conduction band over 4 eV originated from the significant contribution of I-5s with a small share of 6s and 5p states of the Cs atom. Notably, the valence band maximum at the Q-point of the Brillouin zone near the $E_F$ is generated due to the influence of the I-5p state. As a result, in the PDOS of the investigated compound, the orbital electrons of the I-5s/5p orbitals are crucial for absorbing light energy and producing conductivity.

### *3.1.3 Electron charge density of $Cs_2BiAgI_6$ compound*

To understand and give a clear explanation of the charge transformation and nature of bonding among constituent atoms of double perovskite $Cs_2AgBiI_6$, the study of charge density mapping is crucial [54,55]. The charge density mapping along the (111) crystallographic plane is illustrated in **Figure 2(c)**, in which the right scale shows the intensity of electron density (blue and red colors denote the highest and lowest intensity of electron density, respectively). The highest intensity of charge density is observed around the Ag atom, whereas it is lowest around the Bi atom. The charge contours of the Cs and I atoms do not overlap, indicating the existence of ionic bonding between these two atoms. A similar phenomenon is also noticed for Ag and I atoms, revealing the ionic nature of Ag-I bonds. On the other hand, the elliptical contour of charges between Bi and I atoms exhibits the presence of strong covalent bonding of Bi-I bonds. Moreover, isoline sharing suggests that a charge transfer between these two atoms may occur. The strong covalent nature of Bi-I bonds mainly originated from the hybridization between the electronic orbital states of Bi-6p and I-5p.

### 3.1.4 *Optical properties of $Cs_2BiAgI_6$ compound*

The optical functions are essential for understanding the behavior of a material when an incident electromagnetic wave interacts with it. Therefore, these functions bear crucial significance for exploring possible uses of material in photovoltaic and optoelectronic instruments. In such cases, the nature of a specific material under incident photon (infrared, visible, and ultraviolet) is necessary to examine. A number of optical processes, namely, dielectric constant $\varepsilon(\omega)$, refractive index $n(\omega)$, absorption $\alpha(\omega)$, conductivity $\sigma(\omega)$, and reflectivity $R(\omega)$ are determined to investigate the response of $Cs_2AgBiI_6$ upon photon energies up to 30 eV. Generally, both the intra-band and inter-band transitions are used to calculate the $\varepsilon(\omega)$. However, the present calculation overlooks the intra-band contributions that are arisen from indirect transitions. This is because of the inclusion of phonon in the indirect intra-band transition and possesses a limited scattering cross-section [76] related to the direct transition. Hence, there is no need to retain the momentum of phonon scattering. It can be seen from **Figure 3(a)** that the real and imaginary parts of the dielectric function are higher at low energy and remarkably reduced in high-energy regions. This matter hints at the useful usage of $Cs_2AgBiI_6$ in microelectronic devices and integrated circuits [14]. The $\varepsilon(\omega)$ is connected to electron excitement that originates predominantly because of interband transitions and exhibits minor peaks at ~ 6 eV due to intraband transitions. The real and imaginary parts of $n(\omega)$ are shown in **Figure 3(b).** The real part falls gradually in a similar fashion as the real part of the dielectric function, whereas the imaginary part slightly exceeds the real part at ~ 9 eV and then again falls below the real part at ~ 16 eV. The real part of the refractive index is relatively large at zero photon energy, and the maximum is found at ~ 2 eV. It suggests the possible device applications of a chosen compound in quantum-dot light-emitting diodes, organic light-emitting diodes, solar cells, and waveguides [54,73].

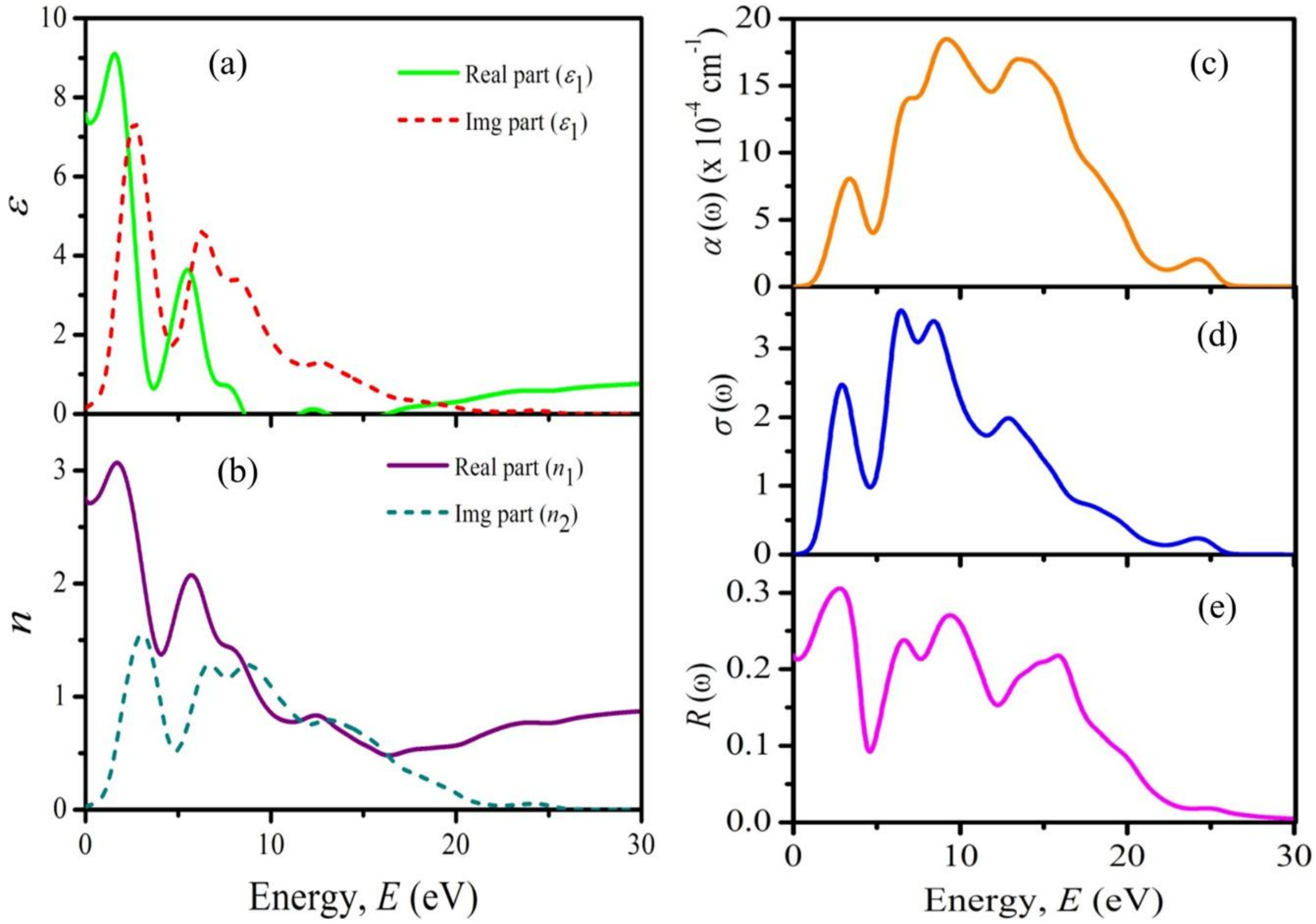

**Figure 3.** Calculated (a) dielectric function, (b) refractive index, (c) absorption, (d) conductivity, and (e) reflectivity of $Cs_2AgBiI_6$.

The $\alpha(\omega)$ presented in **Figure 3(c)** manifests the spectra around 0-26 eV owing to the transitions between energy bands. Importantly, it starts from slightly above zero energy, which reconfirms the small band gap of $Cs_2AgBiI_6$. The $\alpha(\omega)$ is relatively higher at lower photon energy, which predicts the studied compound's potential use in solar panels. Due to the semiconducting Nature of $Cs_2AgBiI_6$, the $\sigma(\omega)$ also does not start at zero energy (**Figure 3(d)**). It starts at ~ 0.84 eV, and the maximum spectrum is attained at 6 eV, which becomes lower with increasing energy. The large conductivity at lower photon energy suggests the conventional applications of $Cs_2AgBiI_6$. It can be noticed from **Figure 3(e)** that the static reflectivity $R(0)$ is ~ 21% of total radiation, which is increased in the high-energy region because of inter-band transitions. Moreover, the $R(\omega)$ exhibits lower values (< 44%) over the studied energy range, indicating significant absorptivity/transmissivity of light, which means that this compound may absorb a significant number of photons and making it suitable for use in optoelectronic devices [54].

## 3.2 Analysis of SCAPS-1D results

### 3.2.1 *Effect of HTL layer*

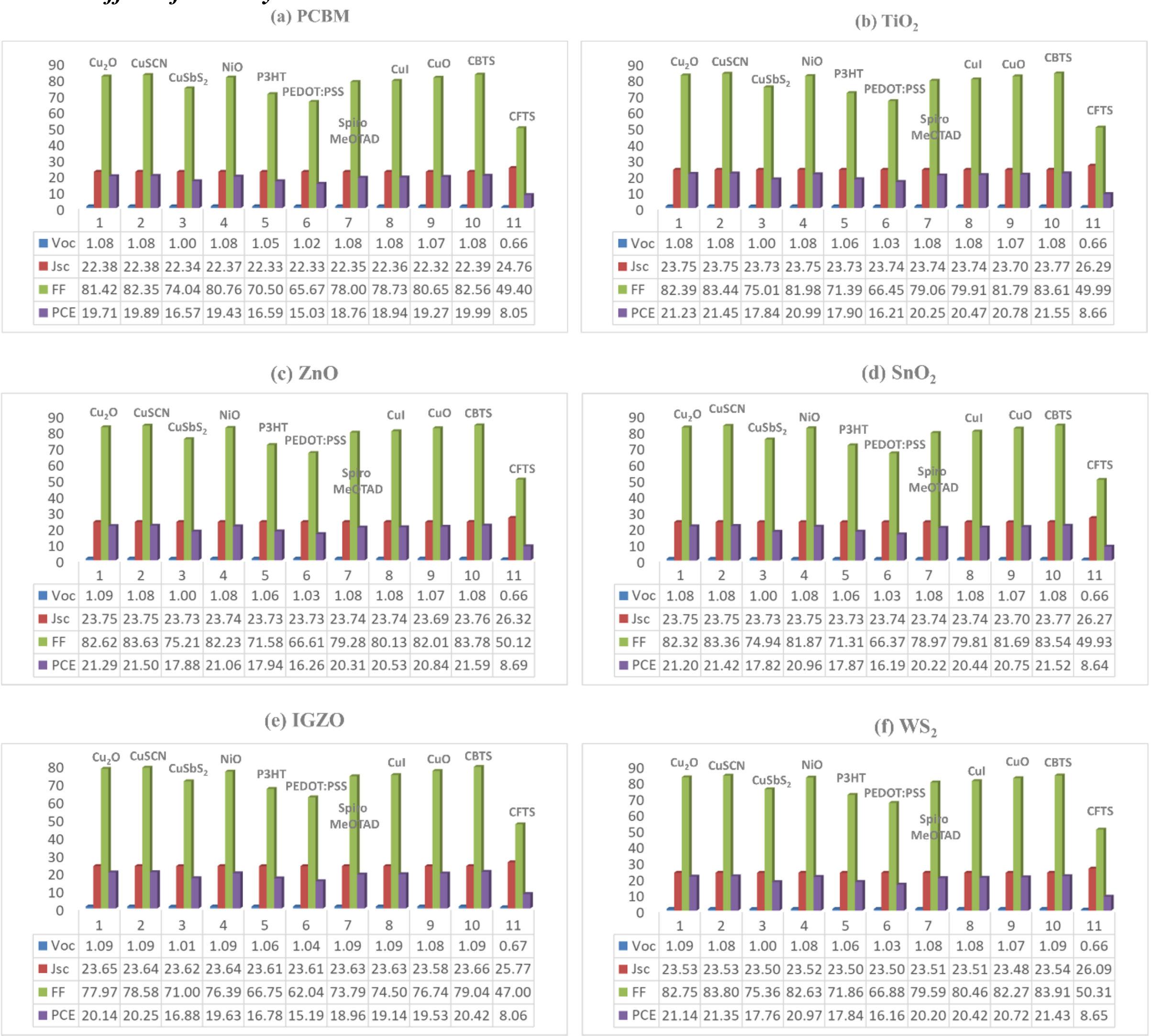

**Figure 4**. Variation of performance parameters, i.e., $V_{OC}$ (V), $J_{SC}$ (mA/cm$^2$), *FF* (%), and PCE (%) of $Cs_2BiAgI_6$ absorber layer-based PSCs device for different HTLs with Au as back metal contact and ETLs of (a) PCBM, (b) $TiO_2$, (c) ZnO, (d) $SnO_2$, (e) IGZO, and (f) $WS_2$.

In the PSCs device configuration, HTL collects holes from the perovskite ($Cs_2BiAgI_6$) and then transfers them to the back metal contact (Au). Twelve types of HTL have been utilized in SCAPS-1D simulation to optimize the device's performance (**Table 2**). The visual depiction of the HTL optimization process is shown in **Figure 4**. **Figure 4(a)** illustrates that PCBM as ETL is optimized with CBTS as HTL compared to the other HTL, while the PCE of this perovskite device configuration is 19.99 %. Besides, $TiO_2$ as ETL showed maximum optimization with CBTS as HTL in **Figure 4(b)**, with a PCE of 21.55 %. **Figure 4(c)** demonstrates that ZnO as ETL with CBTS as HTL exhibited more optimization than other HTL, with a PCE of 21.59%. Similarly, $SnO_2$, IGZO, and $WS_2$ as ETL with CBTS as HTL demonstrated the highest optimization compared to other HTLs. The PCEs of these device configurations are 21.52 %, 20.42 %, and 21.43 % respectively (**Figures 4 (d), (e),** and **(f)**). And the $C_{60}$ and $CeO_2$ as ETL with CBTS as HTL showed the best optimization than other HTLs, with PCE of 17.47% and 14.44%, respectively. Compared to the other HTL presented in this study, CBTS exhibited the highest performance with different ETL. Therefore, CBTS is suitable for HTL optimization among numerous simulated device configurations.

### *3.2.2 ETL Optimization*

In the PSCs device configuration, ETL removes the electron from the perovskite materials and then transfers the electron to the ITO. ETL also prevent electron in ITO from recombining with the absorber layer's holes [77]. Simulating each ETL with different HTL in the $Cs_2BiAgI_6$ perovskite absorber optimizes the performance of device configuration (**Tables 1** and **2**). In this study, we used electron transport materials (ETMs) such as $TiO_2$, PCBM, ZnO, $C_{60}$, IGZO, $SnO_2$, $WS_2$, and $CeO_2$ with different electron transport materials (HTMs). After simulating all possible combinations between the ETM layer and HTM Layer with $Cs_2BiAgI_6$ absorber and Au metal contact, all ETL showed maximum optimization with CBTS as HTL. As a result, eight sets of device configurations from ninety-six combinations exhibited maximum optimized performance and efficiency of device configurations. While band alignment of ZnO (3.3 eV), $TiO_2$ (3.2 eV), and $SnO_2$ (3.6 eV) ETL was significantly higher in comparison with other ETLs which showed ≥ ~21.5% efficiency with CBTS HTL (**Figure 5**). On the contrary, $CeO_2$ (3.5 eV) band alignment isn't convenient with the $Cs_2BiAgI_6$ absorber and CBTS HTL which performed an efficiency of 14.44%. **Table 4** illustrates the performance parameters $V_{OC}$, $J_{SC}$, *FF*, and PCE, which exhibited good agreement value in these eight device configurations.

**Table 4**. Optimized performance parameters of the best combination for each of the ETLs and HTLs

| Optimized Device | Cell thickness (μm) | $V_{OC}$ (V) | $J_{SC}$ (mA/cm$^2$) | *FF* (%) | PCE (%) |
|---|---|---|---|---|---|
| ITO/PCBM/$Cs_2BiAgI_6$/CBTS/Au | 0.5/0.05/0.8/0.1/Au | 1.081 | 22.39 | 82.56 | 19.99 |
| ITO/$TiO_2$/ $Cs_2BiAgI_6$/CBTS/Au | 0.5/0.03/0.8/0.1/Au | 1.084 | 23.80 | 83.61 | 21.55 |
| ITO/ZnO/$Cs_2BiAgI_6$/CBTS/Au | 0.5/0.05/0.8/0.1/Au | 1.085 | 23.76 | 83.78 | 21.59 |
| ITO/$C_{60}$/$Cs_2BiAgI_6$/CBTS/Au | 0.5/0.05/0.8/0.1/Au | 1.077 | 19.62 | 82.65 | 17.47 |
| ITO/IGZO/$Cs_2BiAgI_6$/CBTS/Au | 0.5/0.03/0.8/0.1/Au | 1.092 | 23.66 | 79.04 | 20.42 |
| ITO/$SnO_2$/$Cs_2BiAgI_6$/CBTS/Au | 0.5/0.1/0.8/0.1/Au | 1.084 | 23.77 | 83.54 | 21.52 |
| ITO/$CeO_2$/$Cs_2BiAgI_6$/CBTS/Au | 0.5/0.1/0.8/0.1/Au | 0.924 | 23.59 | 66.21 | 14.44 |
| ITO/$WS_2$/$Cs_2BiAgI_6$/CBTS/Au | 0.5/0.1/0.8/0.1/Au | 1.085 | 23.54 | 83.91 | 21.43 |

### *3.2.3 Band diagram*

The energy band diagram utilizes each ETL with a $Cs_2BiAgI_6$ absorbing layer and CBTS as HTL influencing valence/conduction band offset (i.e., valence band difference between HTL and the absorber layer and conduction band between ETL and absorber layer). The alignment of the energy levels has a significant impact on the efficiency and performance of the PSCs. In the PSCs, photo-generated electrons are injected into the ETL conduction band, and at the same time, holes are transported to the HTL. Afterward, electrons and holes are collected at their respective front (ITO) and back contact metal (Au) correspondingly. The energy band mismatch at the ETL/$Cs_2BiAgI_6$ and the $Cs_2BiAgI_6$/HTL interface significantly impacts the device's performance parameters. The nature of the interface controls the effect of interfacial recombination. Therefore, carefully tuning the electronic properties of ETL and HTL materials became a vital task. Thus, the electron affinity of ETL should be higher than that of $Cs_2BiAgI_6$ to extract the electron safely at the ETL/$Cs_2BiAgI_6$ interface, and the ionization energy of HTL should be lower than that of $Cs_2BiAgI_6$ to extract the holes at the $Cs_2BiAgI_6$/HTL interface. **Figure 5** shows that the Fermi level, which is near the conduction band, enters into the conduction band in the case of the best eight sets of $Cs_2BiAgI_6$-based device configuration. The PCBM as ETM layer and CBTS as HTM layer associated with $Cs_2BiAgI_6$-based device configuration showed that the Fermi level crossed the conduction band, as shown in **Figure 5(a)**. **Figures 5(c)-(h)** exhibited the others of $Cs_2BiAgI_6$-based perovskite device configurations where $TiO_2$, ZnO, $C_{60}$, IGZO, $SnO_2$, $WS_2$, and $CeO_2$ as ETL and CBTS as HTL showed the same pattern and degenerate semiconductors nature as like **Figure 5(a)**.

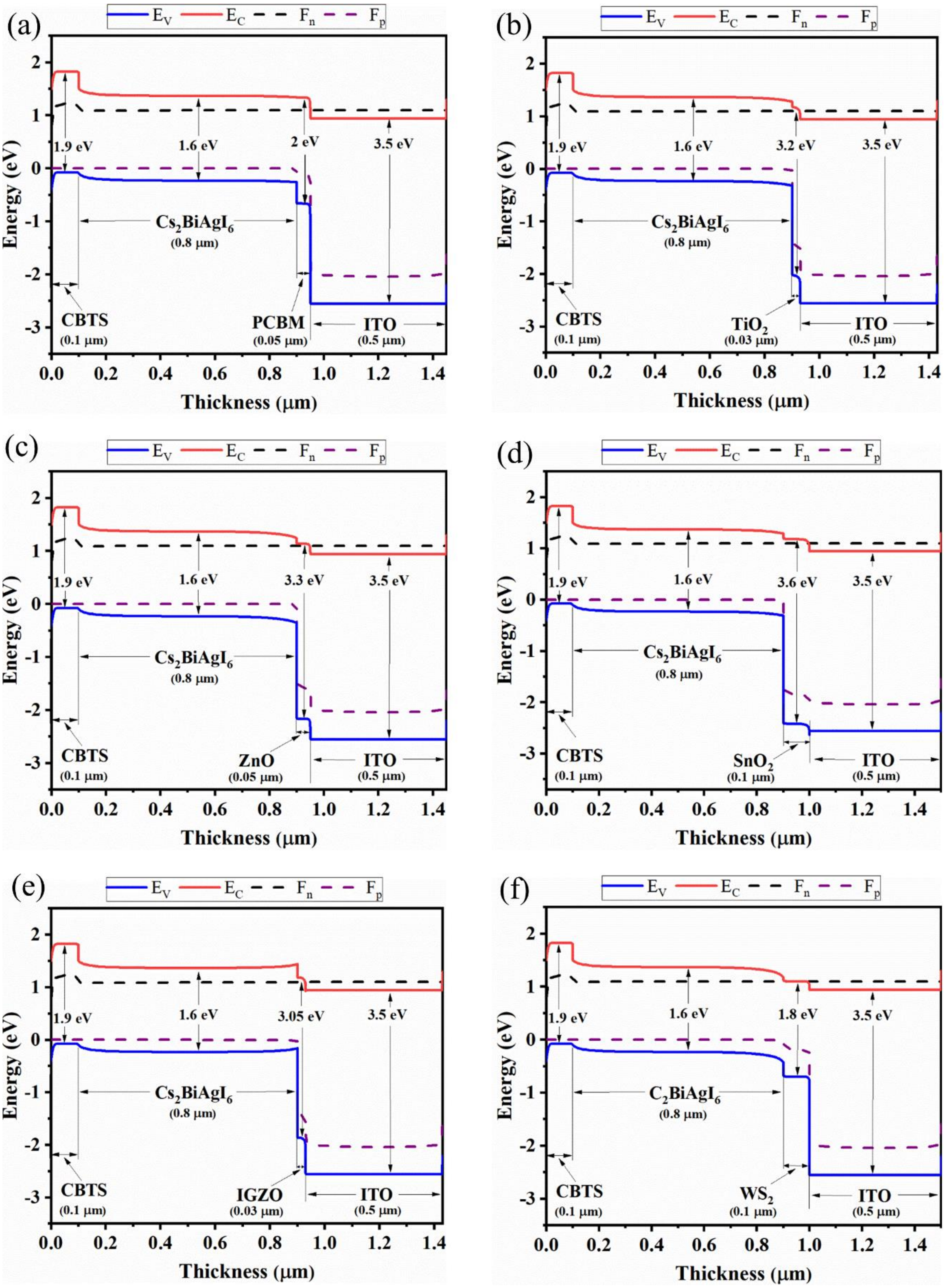

**Figure 5.** Energy band diagram of solar cells structure with different ETMs of (a) PCBM, (b) $TiO_2$, (c) ZnO, (d) $SnO_2$, (e) IGZO, and (f) $WS_2$.

### *3.2.4 Effect of absorber and ETL thickness on cell performance*

The effect of the absorber layers and different ETL layer thicknesses of desired eight double perovskites solar cell structures are investigated through contour plot mapping to evaluate the solar cell's performance parameters. An appropriate absorber and ETL selection of any solar cell is the primary and crucial element in achieving high-performance solar cells. Moreover, the appropriate thickness of the absorber and ETL helps to capture the light and collect holes from the absorber layer, accordingly. **Figure 6** represents the effect of ETL and absorber layer thickness on the $V_{OC}$ parameter of the eight most efficient solar cell structures enlisted in **Table 4**. From **Figure 6(a)**, it is observed that ETL thickness is 0.1 $\mu m$ and absorber thickness is in the range of 0.4 $\mu m$ in the case of ETL as a PCBM-associated structure. In addition, simultaneously increasing ETL thickness with absorber layer thickness can cause lower $V_{OC}$ **(Figure 6(a))**. For the $TiO_2$ and ZnO as ETL-associated structures, the thickness of the absorber layer remains constant with increasing ETL layer thickness **(Figures 6(b)** and **(c))**. The highest value of $V_{OC}$ for $TiO_2$ and ZnO as ETL-associated structures are exhibited at around 0.3 $\mu m$ and 0.5 $\mu m$ absorbers layer thickness regardless of ETL thickness **(Figures 6(b)** and **(c))**. For the $C_{60}$ as ETL associated structure, the highest $V_{OC}$ is exhibited when absorber thickness is around 0.4 $\mu m$, and ETL thickness is about 0.1 $\mu m$ **(Figure 6(d))**. On the other hand, for IGZO as ETL, the higher $V_{OC}$ value is shown when the absorber layer is < 0.4 $\mu m$ with regardless of the ETL layer thickness (**Figure 6(e))**. The $SnO_2$ as ETL-associated solar cell structure shows $V_{OC}$ at a similar range at a particular absorber layer thickness, which is 0.5 to 0.6 $\mu m$ (**Figure 6(f))**. The $V_{OC}$ values of $CeO_2$ as an ETL-associated solar cell structure are shown between 0.1 to 0.2 $\mu m$ ETL thickness and 0.4 to 0.5 $\mu m$ absorber layer thickness (**Figure 6(g))**. According to **Figure 6(h)**, $WS_2$ as an ETL-associated solar structure shows a higher $V_{OC}$ at around 0.03 to 0.25 $\mu m$ of ETL thickness and 0.3 to 0.4 $\mu m$ absorber thickness. In conclusion, the larger thickness of the absorber layer might be the cause of lower $V_{OC}$ for most of the studied solar structures, as observed in **Figure 6.** This fact is explained by the increase in carrier recombination rate in the presence of a thick absorber layer, causing a rise in saturation current larger than photocurrent.

**Figure 7** represents the effect of variation of ETL and absorber layer thickness on the *Jsc* Parameter of eight desired perovskite solar cells. For PCBM as an ETL-associated solar cell, The maximum $J_{SC}$ (22.08 to 24 mA/cm$^2$) value is exhibited when the ETL thickness is 0.03 to 0.1 μm and the absorber thickness is 0.6 to 1.3 μm (**Figure 7(a))**. The almost similar pattern we have observed for ETL $C_{60}$ – associated solar cells, where the higher $J_{SC}$ (19.74 to 22.10 mA/cm$^2$) value is observed when absorber thicknesses are from 0.6 to 1.3 μm and ETL thicknesses are from ≤ 0.1 μm (**Figure 7(d))**. It is observed that the lower ETL thickness and higher absorber thickness can cause higher $J_{SC}$. Higher absorber thickness permits more light collection and hence increases the generation rate, giving a high $J_{SC}$ value. Thin ETL can boost the current via decreasing electron-hole pair recombination which reduces the effect of series resistance. Besides, minimizing the ETL thickness prevents the formation of wider pinholes and rough terrain surfaces, which can severely impede the $J_{SC}$, $V_{OC}$, and therefore efficiency. The maximum $J_{SC}$ value for $TiO_2$ as ETL and ZnO as ETL-associated solar structures are shown to be the same when the absorber layer thicknesses are 0.8 to 1.3 μm, and ETL thickness doesn't make any significant effect during the variation (**Figures 7(b)** and **(c))**. From **Figure 7(e)**, we observe that IGZO as ETL associated solar cell structure shows a maximum $J_{SC}$ pattern, which is almost 24 mA/cm$^2$ when absorber thickness is 0.8 to 1.3 μm, and ETL thickness is 0.03 to 0.3 μm. The $SnO_2$ as an ETL-associated solar structure shows a moderate $J_{SC}$ pattern range with the absorber and ETL thickness variation **(Figure 7(f))**. In the case of $CeO_2$ as ETL, the maximum $J_{SC}$ is 23.58 to 24.20 mA/cm$^2$ when absorber thickness is about 0.8 to 1.3 μm, regardless of ETL thickness (**Figure 7(g))**. Finally, $WS_2$ as an ETL-associated solar cell shows the highest $J_{SC}$ (24.50 mA/cm$^2$) pattern when absorber thickness is 0.7 to 1.3 μm and ETL thickness is 0.03 to 0.2 μm (**Figure 7(g)**).

The effect on *FF* values for the variation of absorber and ETL layer thickness is shown in **Figure 8**. For PCBM as an ETL-associated solar structure, it is observed that *FF* values are decreased with increased absorber and ETL layer thickness **(Figure 8(a)).** This fact is due to the increased series resistance. The lowest *FF* is observed at around 79.16 %, as shown in **Figure 8(a)**. A similar pattern is observed for $TiO_2$, ZnO, $C_{60}$, IGZO, and $WS_2$-associated solar structures **(Figures 8(b)-(e),** and **(h))**. However, **Figure 8(f)** shows that when $SnO_2$ is an ETL, the *FF* values are observed from 31 to 50%. In contrast, the *FF* values for $CeO_2$ as ETL-associated solar cells are increased with decreased absorber thickness (**Figure 8(g))**. An anomaly of the *FF* pattern is observed for different ETL layer thicknesses (around 0.03-0.5 μm) with different absorber layers **(Figure 8(g))**.

**Figures 9(a) and (d)** show that PCBM and $C_{60}$, as ETL-associated solar cells exhibit the same pattern with increasing absorber and ETL layer thickness. The highest PCE of PCBM and $C_{60}$ as ETL is 20.90% and 19.30% when the ETL thickness is less than 0.7 μm and the absorber layer thickness is 0.4 to 1.3 μm. The $TiO_2$ and ZnO as ETL-associated solar cells show the same pattern for PCE with increasing ETL and absorber layer thickness **(Figures 9(b)** and **(c))**. The highest PCE is observed when the absorber layer thickness is more than 0.6 μm regardless of ETL thickness, as shown in **Figures 9(b)** and **(c)**. For IGZO as ETL, the maximum PCE (20.40 %) is observed when the absorber layer thickness is 0.5 to 1.1 μm and the ETL layer thickness is 0.03 to 0.1 μm **(Figure 9(e))**. The PCE pattern is almost moderate in the case of $SnO_2$ as an ETL-associated solar structure **(Figure 9(f))**. For $CeO_2$ as ETL, the highest PCE is observed when absorber thickness is 0.5 to 0.7 μm, and ETL thickness does not significantly impact the PCE pattern, as shown in **Figure 9(g)**. Finally, in the range of 0.5 to 1.5 μm absorber and less than 0.15 μm ETL thickness, the PCE is highest for $WS_2$ as an ETL-associated solar structure (**Figure 9(h)).**

The changing trend of $V_{OC}$, $J_{SC}$, $FF$, and PCE of perovskite solar cells with different ETL is very distinct due to several reasons. The absorption coefficient significantly affects the PV parameters like $V_{OC}$, $J_{SC}$, $FF$, and PCE and is directly linked with the bandgap of the ETLs. It also affects the coupling of incident photons in the underlying $Cs_2BiAgI_6$ absorber layer. The conduction band offset (CBO) which depends on the workfunction (affinity + fermi energy level) difference between ETL and absorber layer may also play a crucial role in the performance of the solar cell. Where there is a preferred electron affinity, it can have a big impact on how well the solar cell works. Energy cliffs with CBO (-) and energy spikes with CBO (+) are produced when the electron affinities of the ETL and absorber layers are different. In other words, if the position of the conduction band (CB) of the ETL is lower than the absorber, the energy Cliff CBO (-) is formed at the ETL/absorber interface without the potential to act as an electron barrier [78]. So, PCBM, $C_{60}$, and $CeO_2$ have no electron barrier on the ETL/absorber interface. On the contrary, if the energy spike-CBO (+) is formed at the ETL/absorber interface it will act as an electron barrier [78]. Therefore, ETLs like $TiO_2$, ZnO, IGZO, $SnO_2$, and $WS_2$ have electron potential barriers at the ETL/absorber interface. Such kind of spike structure developed at the ETL/absorber layer interface, which can operate as a barrier for photo-generated electron flow towards the edge of the ETL/absorber, has the advantage of enhancing the photo generation of free charge carriers.

Additionally, by doing this, the rate of recombination at the contact will be reduced, improving PV performance. The primary recombination process occurs at the interface when the activation energy for carrier recombination ($E_A$) is lower than the bandgap of the absorber layer, as opposed to the creation of cliff structures between ETL and perovskite. Additionally, the development of "cliff-type" band alignment will lower the resistance to electron transfer, resulting in a fall in $E_A$, which directly influences the value of $Voc$ and, in turn, the PV parameters like $J_{SC}$, $FF$, and PCE of the solar cell [79]. The impact of CBO on $V_{OC}$ can be described according to **Eqs. 8-9** where $V_{OC}$ denotes open circuit voltage, $E_A$ denotes the activation energy, $n$ denotes the diode ideality factor, K denotes Boltzmann constant, T denotes temperature, $J_{00}$ denotes current pre factor and $J_{SC}$ denotes short circuit current density:

$$V_{OC} = \frac{E_a}{q} - \frac{nKT}{q}\ln\frac{J_{00}}{J_{SC}} \tag{8}$$

$$E_A = E_g - \text{CBO} \tag{9}$$

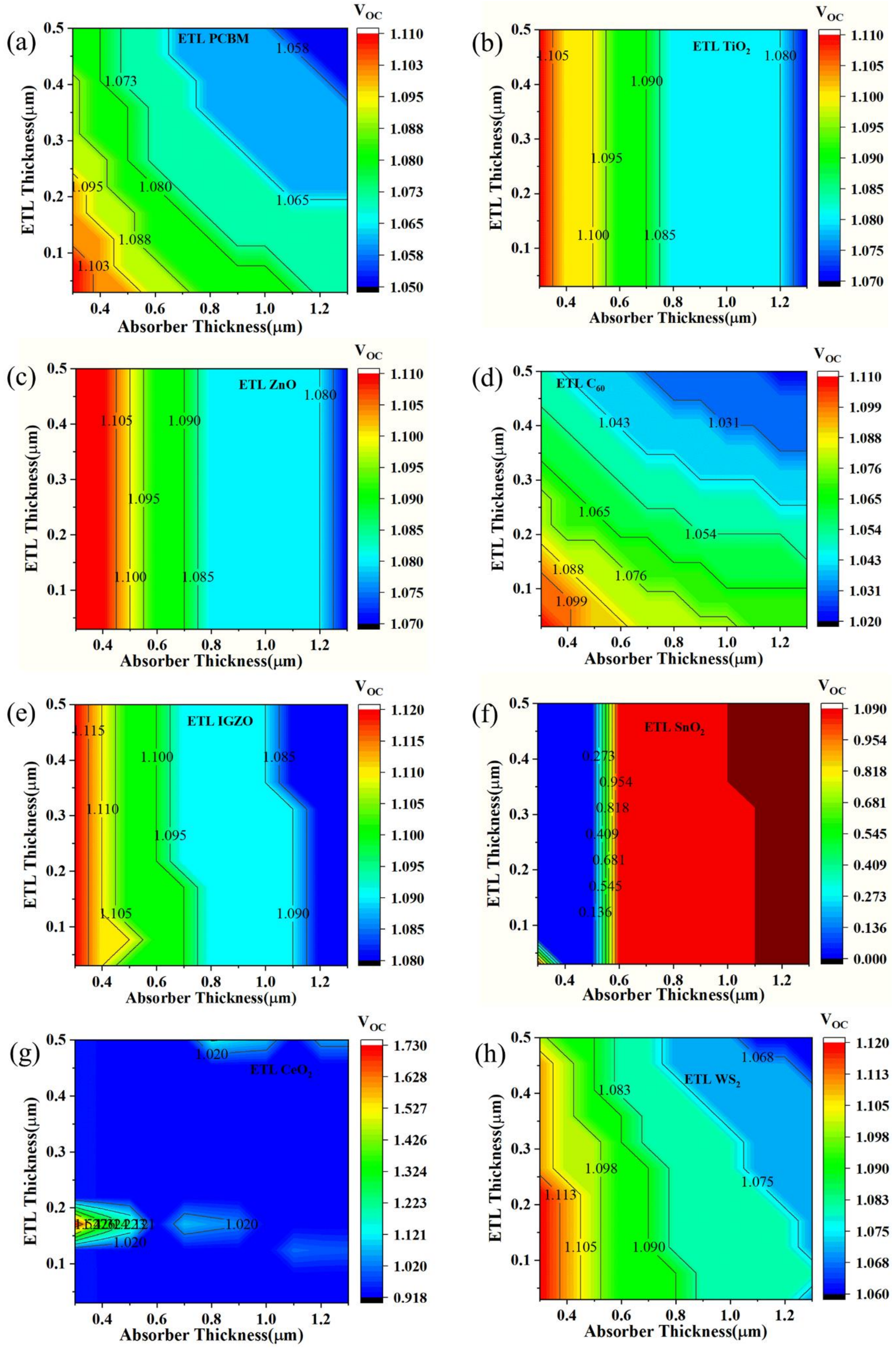

**Figure 6**. Contour mapping of $V_{OC}$ (V) when ETL as (a) PCBM, (b) $TiO_2$, (c) ZnO, (d) $C_{60}$, (e) IGZO, (f) $SnO_2$, (g) $CeO_2$, and (h)$WS_2$.

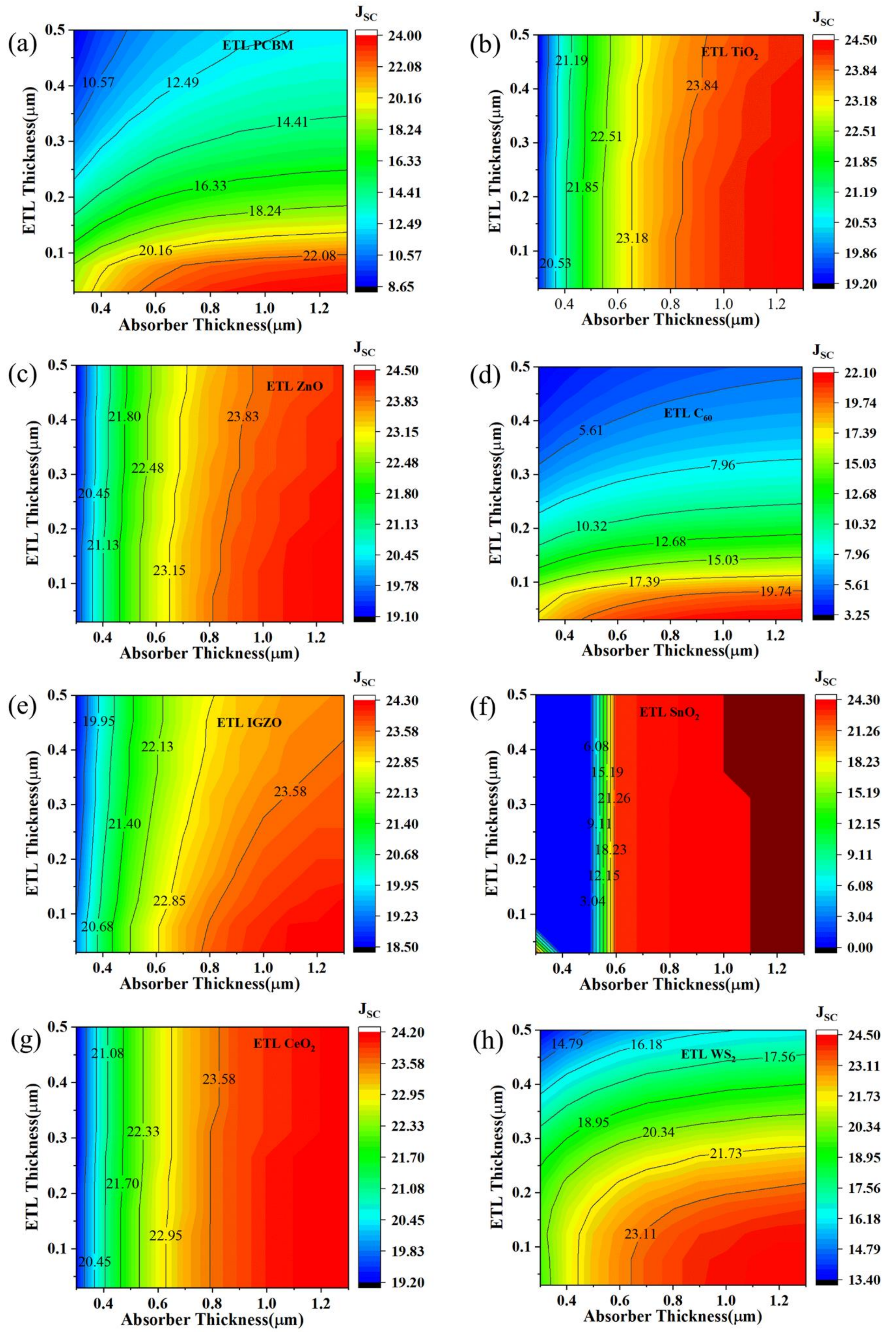

**Figure 7**. Contour mapping of $J_{SC}$ (mA/cm$^2$) when ETL as (a) PCBM, (b) $TiO_2$, (c) ZnO, (d) $C_{60}$, (e) IGZO, (f) $SnO_2$, (g) $CeO_2$, and (h)$WS_2$.

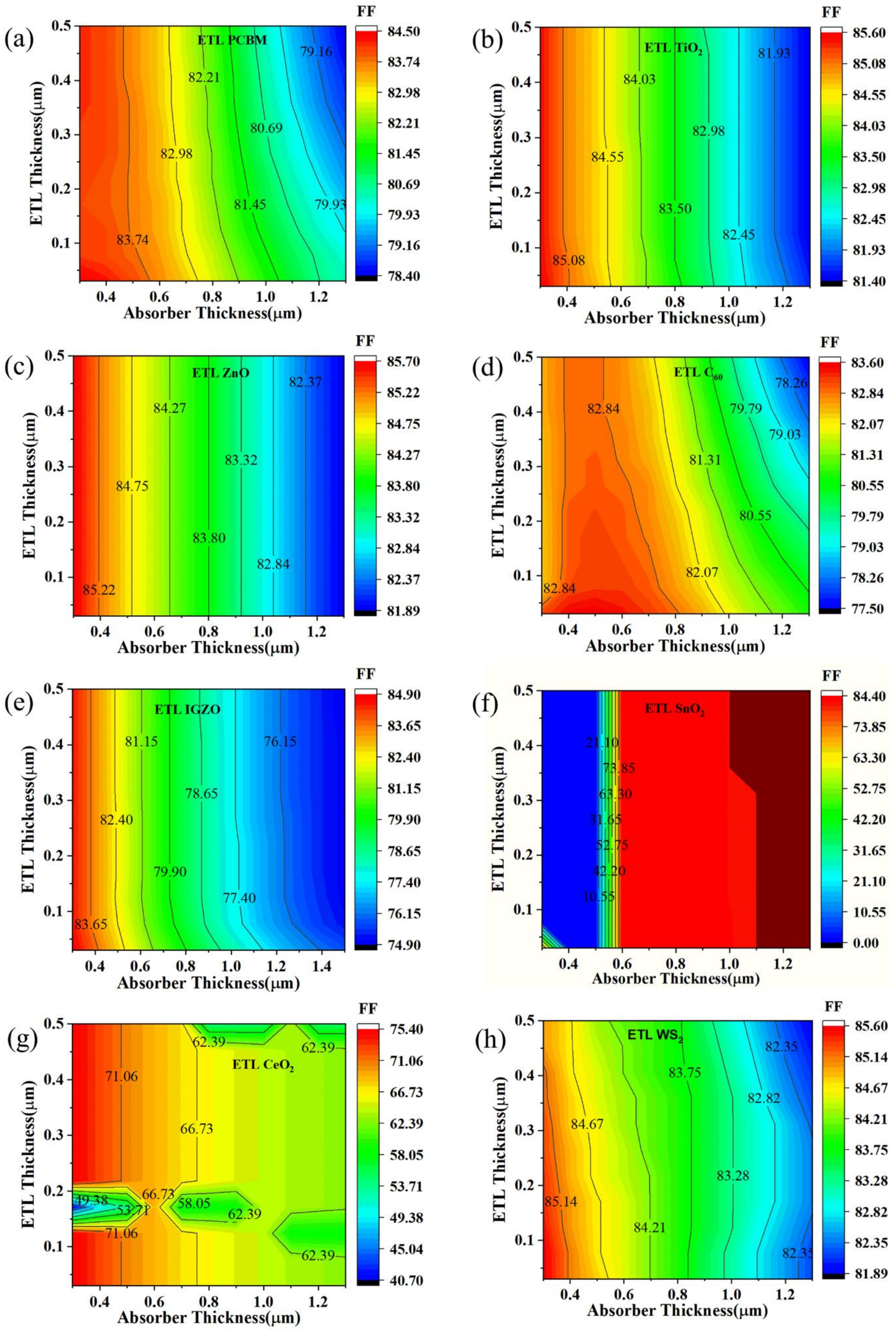

**Figure 8**. Contour mapping of *FF* (%) when ETL as (a) PCBM, (b) $TiO_2$, (c) ZnO, (d) $C_{60}$, (e) IGZO, (f) $SnO_2$, (g) $CeO_2$, and (h)$WS_2$.

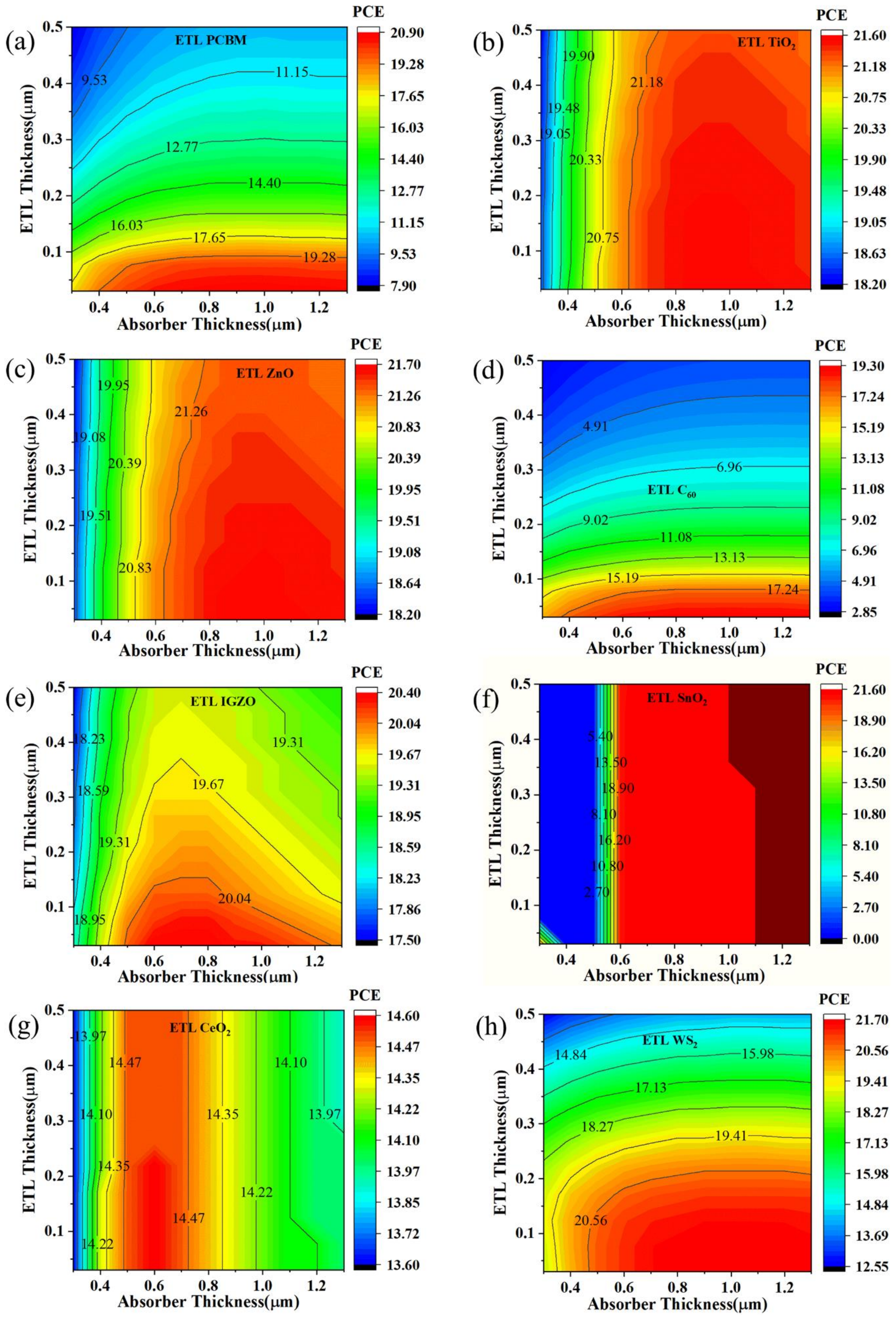

**Figure 9**. Contour mapping of PCE (%) when ETL as (a) PCBM, (b) $TiO_2$, (c) ZnO, (d) $C_{60}$, (e) IGZO, (f) $SnO_2$, (g) $CeO_2$, and (h)$WS_2$.

### 3.2.5 *Effect of series resistance*

The performance of solar cells is significantly impacted by the shunt ($R_{Sh}$) and series ($R_s$) resistances, which are produced mainly by connections among the solar cell layers, right and left side metal contacts, and manufacturing flaws [14]. According to **Figure 10,** the effect of $R_S$ varied from 0-6 Ω-cm$^2$ while the shunt resistance was constant at $10^5$ Ω-cm$^2$ in the case of eight double perovskites (ITO/ETL/$Cs_2BiAgI_6$/CBTS/Au) devices. During the $R_S$ variation, PCE was decreased for all eight $Cs_2BiAgI_6$ perovskite device structures. The PCE value of $SnO_2$, ZnO, $WS_2$, and $TiO_2$ as ETL-based $Cs_2BiAgI_6$ perovskite device structure decreased from around 21.60 to 19.30 %. In contrast, PCBM and IGZO as ETL-based $Cs_2BiAgI_6$ perovskite device structure decreased from about 20 to 19% with increasing $R_S$. And $C_{60}$ and $CeO_2$ ETL-associated solar cell structures, showed almost 17.5 to 16% and 14.5 to 12% PCE, respectively with the increase of $R_S$. *FF* value also decreased with the increase of $R_S$ while the *FF* value of all ETL-associated solar cells exhibited a higher value except IGZO and $CeO_2$ as ETL. The $J_{SC}$ and $V_{OC}$ performance showed the constant value for all ETL-associated device configurations with increasing $R_S$. That means $R_S$ variation does not impact the $J_{SC}$ and $V_{OC}$ parameters for the studied device configurations.

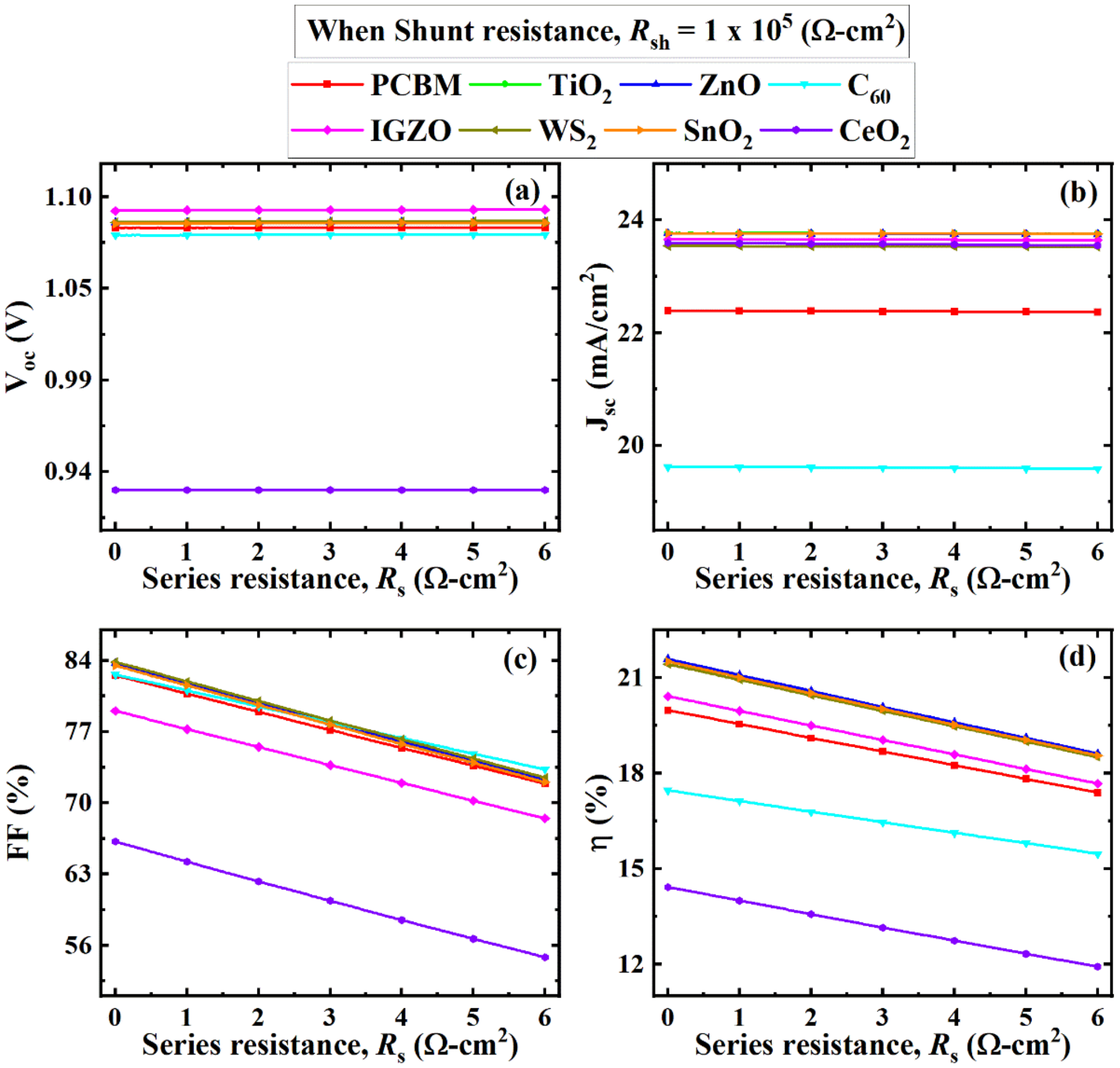

**Figure 10.** Effect of $R_S$ on performance parameters of (ITO/ETL/$Cs_2BiAgI_6$/CBTS/Au, ETL = PCBM, $TiO_2$, ZnO, $C_{60}$, IGZO, $WS_2$, $SnO_2$, and $CeO_2$) double PSCs.

From **Figure 10(b)** is seen that the current remains constant with the larger series resistance. $CH_3NH_3SnI_3$-based study the performance was analyzed by varying $R_S$ from 0 to 6 Ω-cm$^2$ while keeping $R_{sh}$ fixed at $10^5$ Ω-cm$^2$ performed a similar trend of results where the current was unaffected by series resistance, reported by Sunny et al. [80]. Another study reported that $FAPbI_3$, $FA_{0.85}Cs_{0.15}PbI_3$, and $FA_{0.85}Cs_{0.15}Pb(I_{0.85}Br_{0.15})_3$-based perovskites, $R_S$ didn't affect much the current for a certain range of 30 Ω-cm$^2$. After this range, $R_S$ of ≥ 30 Ω-cm$^2$ currents for those particular devices tended to decline [81]. $R_S$ is the sum of resistances between various terminals like absorber, ETL, and HTL as well as the front and back contacts of the cell which doesn't affect current up to a certain range.

So, it could be concluded that the current of the $Cs_2BiAgI_6$ perovskite was affected by various studied ETLs after the larger series of resistance above 30 Ω-cm², which is in line with previous studies.

### *3.2.6 Effect of shunt resistance*

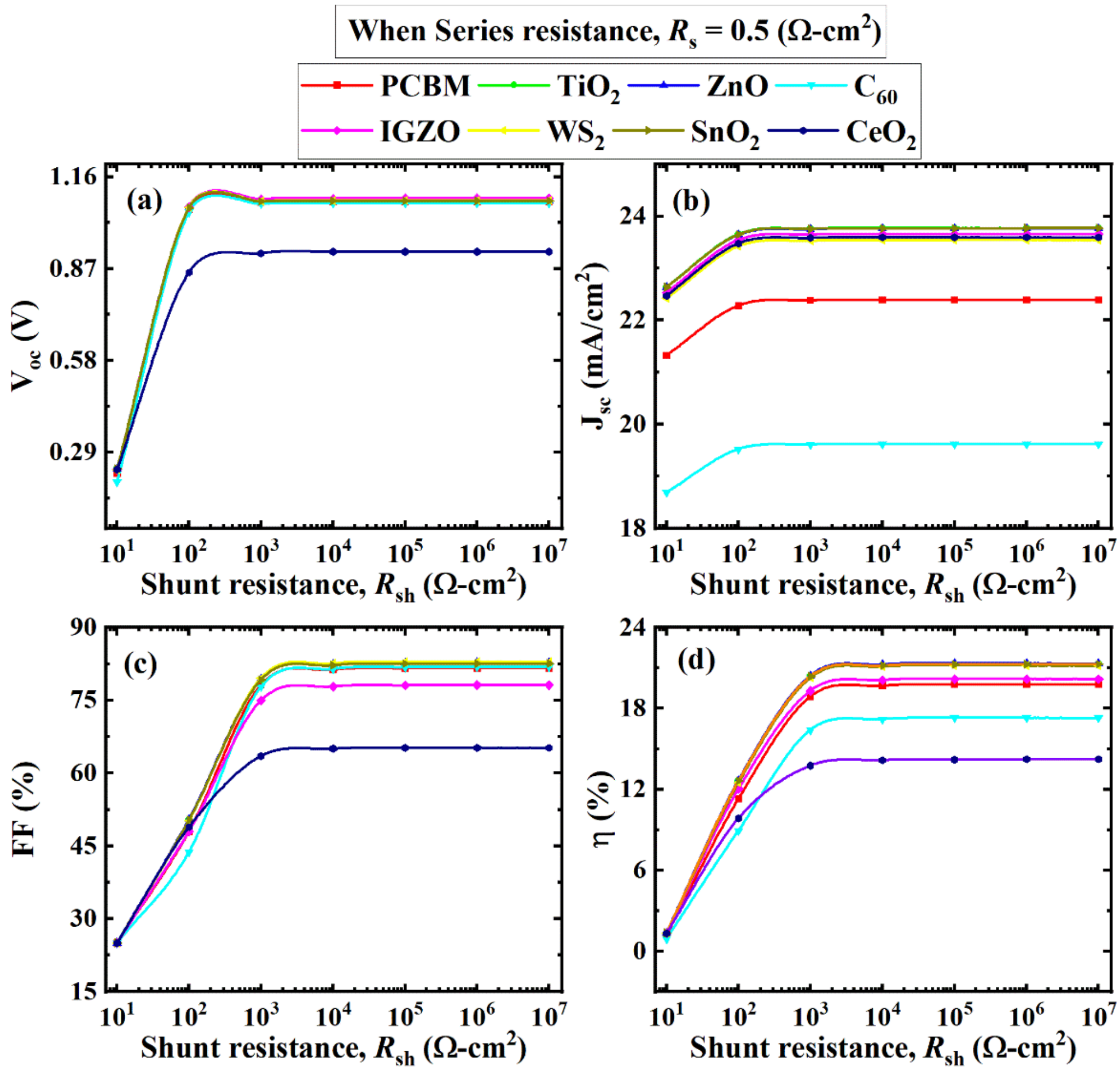

**Figure 11.** Effect of $R_{Sh}$ on performance parameters of (ITO/ETL/$Cs_2BiAgI_6$/CBTS/Au, ETL= PCBM, $TiO_2$, ZnO, $C_{60}$, IGZO, $WS_2$, $SnO_2$, and $CeO_2$) double PSCs.

Internal resistances, interface barriers, charge-collecting interlayers, and metal-based electrodes, are the source of $R_S$ in PSCs, whereas leakage channels, like pinholes in the photoactive layer and recombination losses, are the source of the $R_{Sh}$ [82]. The Shockley equation is expressed as **Eqs. (10)** and **(11)**, describe the expected behavior of the *J-V* characteristic of a solar cell during ideal one-sun illumination conditions [83].

$$J_{SC} = J_{PH} - J_o[\exp(\frac{q_e(V-JR_S)}{nkT_e}) - 1] - \frac{V-JR_S}{R_{sh}} \quad (10)$$

$$V_{OC} = (\frac{nkT_e}{q_e})\ln\{\frac{J_{PH}}{J_0}(1 - \frac{VOC}{J_{PH}R_{Sh}})\} \quad (11)$$

Where, $q_e$ is the elementary charge, $J_{PH}$ is the photocurrent density, $J_o$ is the density of the reverse bias saturation current., $R_s$ is the series resistance, $R_{Sh}$ is the shunt resistance, n is the diode ideality factor, k is the Boltzmann constant (1.38 x $10^{23}$ $JK^{-1}$), and $T_e$ is the ambient temperature (298 K). Also, from **Eqs. (8)** and **(9)**, it is seen that $J_{SC}$ and $V_{OC}$ show negative inverse relation with $R_{Sh}$. That means with increasing $R_{Sh}$, $J_{SC,}$ and $V_{OC}$'s increment happens.

The effect of $V_{OC}$, $J_{SC}$, *FF*, and PCE values with $R_{Sh}$ variation is visually represented in **Figure 11,** where $R_{Sh}$ varied from $10^1$ to $10^7$ *Ω*-cm² for all eight optimum solar cell structures. The $V_{OC}$, $J_{SC}$, *FF*, and PCE values showed a similar pattern with increasing shunt resistance $R_{Sh}$. All performance parameters increased rapidly from $10^1$ *Ω*-cm² to $10^3$ *Ω*-cm² and then maintained the constant value with increasing $R_{Sh}$. The $CeO_2$ as ETL associated structure showed the lowest value where $CeO_2$ initially increased and remained constant for $V_{OC}$, *FF*, and PCE from 0.25 V, 24%, and 2.5% to 0.9 V, 65%, and 13%, respectively, as shown in **Figure 11**. In **Figure 11(b)**, The $C_{60}$ exhibited the lowest value for $J_{SC}$, from 18.5 to 19.5 (mA/cm$^2$). The ZnO, PCBM, $TiO_2$, IGZO, $SnO_2$, $C_{60}$, and $WS_2$ as ETL-associated solar cell structures showed an almost similar and constant value of $V_{OC}$, which is 1.1 V after $R_{Sh}$ of $10^3$ $\Omega$-cm$^2$. The $J_{SC}$ value of $TiO_2$, ZnO, IGZO, $SnO_2$, $WS_2$, and $CeO_2$ as ETL-associated solar cell structure showed 21.0 (mA/cm$^2$) initially and then increased to 23.5 (mA/cm$^2$) and remained constant (**Figure 11(b)**).

### *3.2.7 Effect of temperature*

One of the most important factors in determining the stability of solar cell performance is having a thorough understanding of solar cell performance at high operating temperatures. Due to the distortion between layers at high temperatures, the performance of the majority of solar cell architectures exhibits instability. Recent studies on perovskites-based optoelectronics have shown improvements in the device's performance stability at high temperatures [84–86]. The temperature range has been adjusted from 275 K to 475 K to investigate the relationship between temperature and the efficiency of the solar cell. **Figures 12(a), (c),** and **(d)** shows PCE, *FF*, and $V_{OC}$ values are decreased with increasing temperature for almost all optimum solar cell structures.

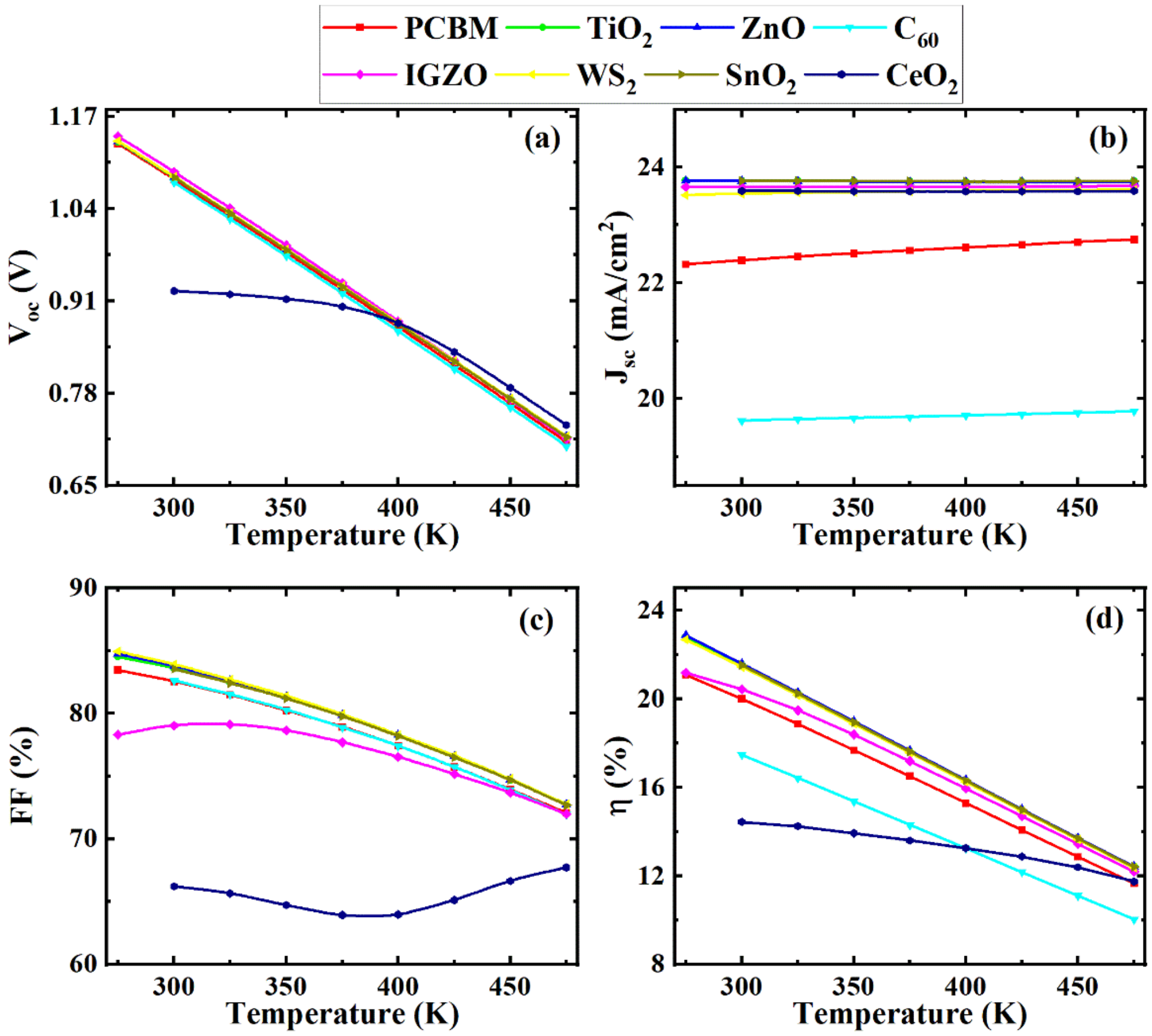

**Figure 12.** Effect of temperatures on performance parameters of (ITO/ETL/$Cs_2BiAgI_6$/CBTS/Au, ETL= PCBM, $TiO_2$, ZnO, $C_{60}$, IGZO, $WS_2$, $SnO_2$, and $CeO_2$) double PSCs.

In contrast, $J_{SC}$ remained almost constant for all optimum device configurations with temperature variation **(Figure 12(b))**. The $C_{60}$ as an ETL-associated solar cell showed a lower $J_{SC}$ of 20 mA/cm$^2$. The PCBM as an ETL-associated structure demonstrated almost 22 mA/cm$^2$ of $J_{SC}$ value, which is higher than the $C_{60}$. The other ETLs with device configurations exhibited around 23.5 mA/cm$^2$ of $J_{SC}$ value with increasing temperature. The $V_{OC}$ value was decreased from 1.16 to 0.70 V with increasing temperature for all studied device configurations except $CeO_2$ as ETL associated device, which decreased from 0.92 to 0.72V (**Figure 12(a)**). The *FF* value decreased from around 85 to 74% for all optimum device configurations except IGZO and $CeO_2$ ETL-associated devices. The PCE decreased for all devices with the increasing temperature. However, the $V_{OC}$ value decreased for all optimum device configurations with increasing temperature due to the inverse relationship between $V_{OC}$, and the reversed saturation current density, $J_0$. The $J_0$ increased at a higher temperature. **Eq. 12** shows the relation between $V_{OC}$ and $J_0$.

$$V_{OC}= \frac{Ak'T_l}{q}[\ln(1+\frac{J_{SC}}{J_0})] \quad (12)$$

Where, $\frac{k'T_l}{q}$ stands for the thermal voltage, and A stands for the ideality factor.

In addition, when the temperature of the PSC rises, the flaws become worse, and the $V_{OC}$ value decrease which is consistence with previous studies [14]. From **Figure 12(b)**, it is evident that the current was affected slightly due to the reduction of bandgap with the increase in temperature. But the range of change is very marginal which seems to be constant during the increment of temperature. And, $CH_3NH_3SnI_3$ perovskite showed a similar kind of trend of change in current with an increase in temperature from 260 K to 473 K [80]. In these circumstances, the bandgap reduction could be less of being double perovskite such kind of a slight current change for temperature change.

### 3.2.8 *Effect of capacitance and Mott-Schottky*

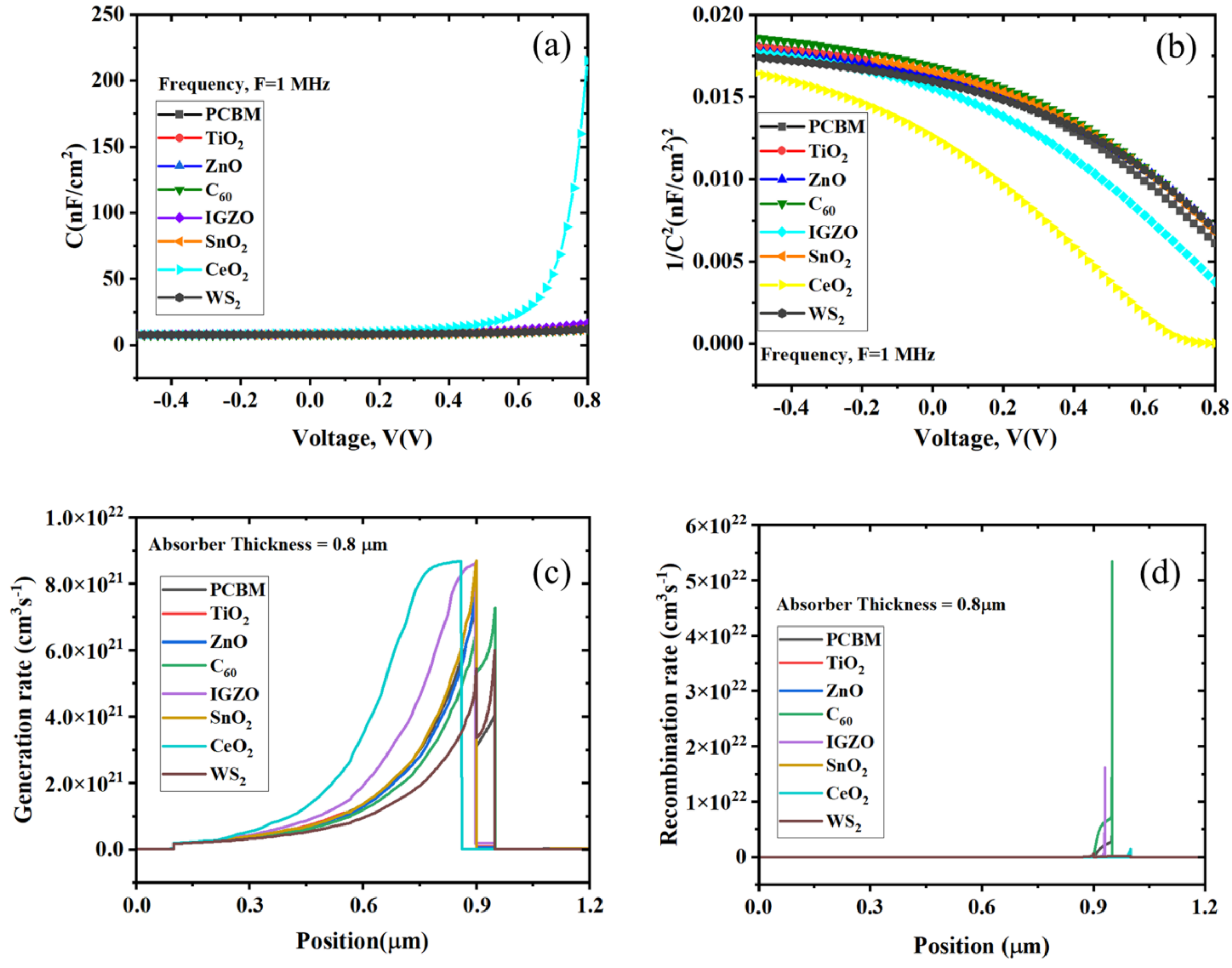

**Figure 13**. (a) Capacitance (C), (b) Mott-Schottky ($1/C^2$) plot, (c) generation rate, and (d) recombination rate for eight studied structures.

**Figures 13(a)** and **(b)** show the Mott-Schottky (M-S) plots and the plots of capacitance per unit area (C) with bias voltage (V), respectively, for eight suitable solar cells. The built-in voltage ($V_{bi}$) and charge carrier density ($N_d$) can be extracted from C-V measurements by using the well-known M-S analysis experimental method. It is used in traditional semiconductor devices that contain p-n and semiconductor/metal junctions with fixed depletion layers and space charge regions. **Eq. 13** yields the junction capacitance per area (C) value.

$$\frac{1}{C^2} = \frac{2\varepsilon_0\varepsilon_r}{qN_d}(\nu_{bi} - V) \tag{13}$$

Here, $\varepsilon_0$ is the vacuum permittivity, $\varepsilon_r$ is the donor's dielectric constant, q is the electronic charge, and $V$ is the applied voltage **(Figure 13(b))** [87,88]. $N_d$ is generated from the gradient of the linear component, and $V_{bi}$ is generated from an extension of the linear part to the voltage axis. The $CeO_2$ ETL-associated device shows voltage-independent capacitance from -0.4 V to 0.5 V, and an exponentially increasing pattern has been shown after 0.5 V **(Figure 13(a))**. Whereas other ETL-associated solar structures represent the independent voltage capacitance it may be because of the saturation of depletion layer capacitance, as seen in **Figure 13(a).** In the case of $CeO_2$ as ETL associated structure $V_{bi}$ shows declining in nature where ZnO, $C_{60}$, PCBM, $SnO_2$, $TiO_2$, and $WS_2$ as ETL associated structure shows a higher voltage.

### *3.2.9 Effect of generation and recombination rate*

**Figures 13(c)** and **(d)** show the generation and recombination rates, respectively. The electron-hole pairs are created during the carrier generation process when an electron is excited from the valence band to the conduction band, producing a hole in the valence band. The emission of electrons and holes brings on carrier generation. To determine the creation of electron-hole pairs G(x), SCAPS-1D employs the incident photon flux $N_{phot}$ ($\lambda$, x). From this photon flux for each position and wavelength, the value of G(x) can be computed through **Eq. 14**.

$$G\,(\lambda, x) = \alpha(\lambda, x).\,N_{phot}\,(\lambda, x) \tag{14}$$

The process of recombination is the opposite of the process of generation, in which the electrons and holes of the conduction band are coupled and annihilated. The charge carrier's lifespan and density determine the solar cell's recombination rate. First, the electron-hole recombination is decreased because of the defect states present within the absorber layer. Then, the energy states are produced, which impacts the electron-hole recombination profile within the solar cell structure. Due to flaws and grain boundaries, the recombination rate distribution is not uniform, as shown in **Figure 13(d)** [54].

### *3.2.10 J-V and QE characteristics*

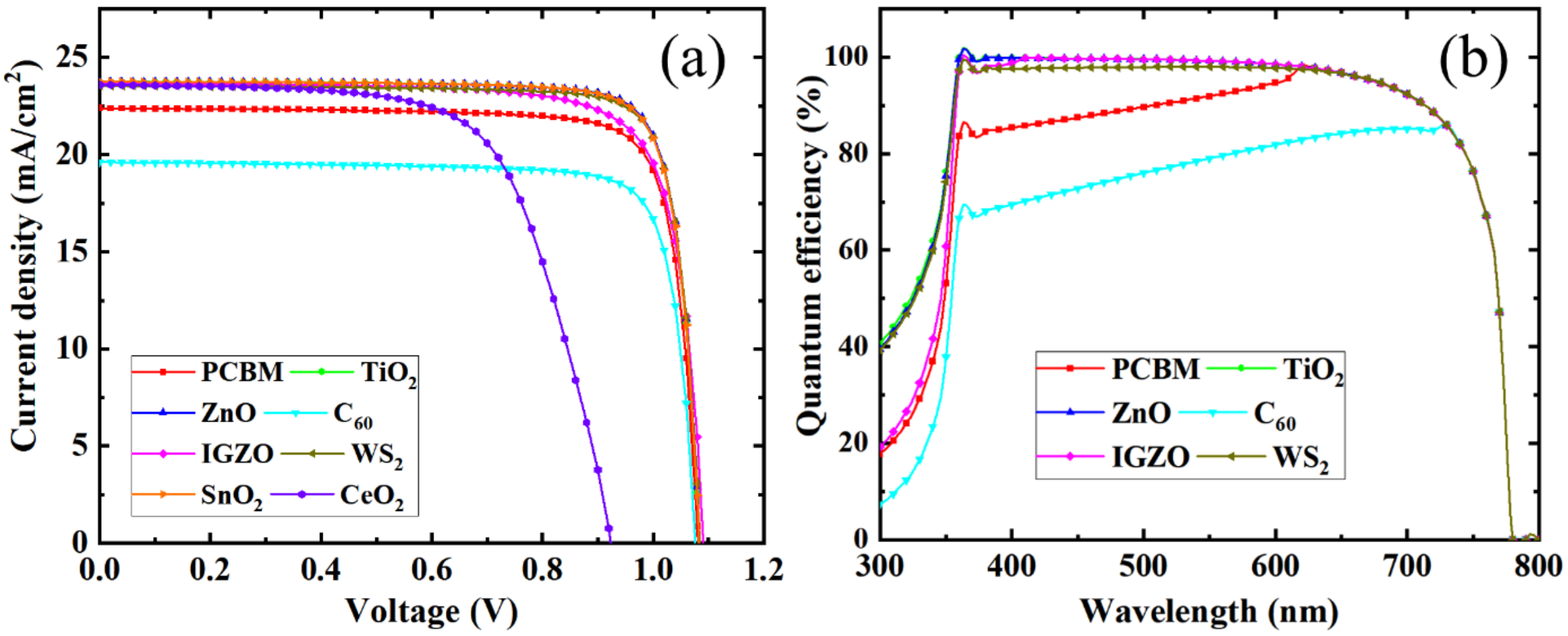

**Figure 14**. (a) *J-V* characteristics and (b) *QE* curve of the double PSCs.

**Figure 14(a)** shows the *J-V* characteristics changing pattern of the studied eight solar cell structures. According to **Figure 14(a)**, it is noticed that $C_{60}$ as an ETL-associated PSC showed about 19 mA/cm$^2$ as current density, while the $V_{OC}$ is more significant than one volt. However, $CeO_2$ as an ETL-associated device exhibited a $J_{SC}$ of about 23 mA/cm$^2$ although the $V_{OC}$ was less than 1.0 V. However, the other ETLs associated structures showed better performance than $C_{60}$ and $CeO_2$ as ETL-associated device structures. The ZnO as ETL-associated structure demonstrated 23 mA/cm$^2$ with unity open circuit voltage density, as shown in **Figure 14(a)**. The defect amount in perovskite films is a critical factor in determining the device's performance since the photoelectrons are produced in these layers. However, the electron-hole recombination behaviors are essential in defining the photovoltaic parameters of the PSC. The *J-V* curve in a perovskite film is shown in **Figure 14(a)** as a function of bulk trap density**.** All photovoltaic parameters are drastically reduced when defect states are present in perovskite films. This is consistent with the observation that significant crystallinity in perovskite reduces charge recombination and enhances performance [44].

**Figure 14(b)** shows the quantum efficiency (*QE*) curve with respect to a different wavelength. The $C_{60}$ as ETL with PSC showed the optimum *QE* of 80% when the wavelength was almost 700 nm. The optimum *QE* of PCBM as an ETL-associated solar cell structure exhibited around 95% at around 620 nm. In addition, the other four ETL-associated PSCs demonstrated the highest *QE* of around 100% at 350 nm. **Figure 14(b)** shows that the six sets of PSCs do not have *QE* above 800 nm. However, because of recombination, where charge carriers cannot pass into an external circuit, the *QE* for most solar cells is decreased. The *QE* is impacted by the same processes that impact collection probability. For instance, altering the front surface might impact carriers produced close to the surface. Free carrier absorption, which lowers *QE* in longer wavelengths, may also be brought on by highly doped front surface layers [89].

## 3.3 Comparison with wxAMPS simulation results and previous work

### *3.3.1 Comparison between SCAPS-1D and wxAMPS results*

A comparison of $Cs_2BiAgI_6$-based double perovskite solar cells is conducted in this section. **Table 5** gives a comparative view of the performance parameters of eight $Cs_2BiAgI_6$ base solar cell structures found by optimizing different ETL and HTL from numerous simulations. PCBM as ETL associated with $Cs_2BiAgI_6$ based perovskite solar structure shows 19.99% of PCE in SCAPS-1D simulator whereas 20.70% of PCE in wxAMPS simulator (**Table 5**). In addition, $TiO_2$ as ETL associated with $Cs_2BiAgI_6$-based solar cells shows 21.55 % of PCE while 23.06% of PCE in the wxAMPS simulator. ZnO as an ETL-associated solar cell structure shows 21.59% PCE when SCAPS-1D is used, whereas wxAMPS shows 22.04% PCE. Similarly, $C_{60}$, IGZO, $SnO_2$, $CeO_2$, and $WS_2$ as ETL-associated $Cs_2BiAgI_6$-based solar cell structures show 17.47%, 20.41%, 21.52%, 14.44%, 21.43% of PCE in SCAPS-1D simulator however in wxAMPS simulator, the PCE values show 21.03%, 22.38%, 21.93%, -, 21.23%. In the case of $V_{OC}$, PCBM as an ETL-associated structure shows 1.08 V in the SCAPS-1D simulator. In contrast, in the wxAMPS simulator, the device structure shows 1.18 V. Like PCBM as ETL associated structure, both simulator shows similar pattern results for $V_{OC}$ for all device configurations. In addition, the performance parameters $J_{SC}$ and *FF* in SCAPS-1D and wxAMPS simulator show the closest pattern for all device structures. Comparing the performance parameters of the eight solar cell structures, SCAPS-1D and wxAMPS simulators show the nearest results except for PCE and $V_{OC}$, where wxAMPS shows a little higher result than SCAPS-1D (**Table 5**). The obtained performance parameters result from both simulators help us to understand the charges transport layers and absorber materials that are suitable for device configuration. Since both simulators give the closest performance results for eight sets of optimum solar device configurations, the presented results are reliable for further investigation which is consistent with previous studies [40,90].

**Table 5**: Comparison between SCAPS 1D and wxAMPS software simulation.

| Device structure | Software | $V_{OC}$ (V) | $J_{SC}$ (mA/cm$^2$) | *FF* (%) | PCE (%) |
|---|---|---|---|---|---|
| ITO/PCBM/$Cs_2BiAgI_6$/CBTS/Au | SCAPS-1D | 1.08 | 22.39 | 82.56 | 19.99 |
| | wxAMPS | 1.18 | 20.86 | 84.21 | 20.70 |
| ITO/$TiO_2$/$Cs_2BiAgI_6$/CBTS/Au | SCAPS-1D | 1.08 | 23.8 | 83.61 | 21.55 |
| | wxAMPS | 1.18 | 23.00 | 84.69 | 23.06 |
| ITO/ZnO/$Cs_2BiAgI_6$/CBTS/Au | SCAPS-1D | 1.085 | 23.76 | 83.78 | 21.59 |
| | wxAMPS | 1.18 | 21.97 | 84.81 | 22.04 |
| ITO/$C_{60}$/$Cs_2BiAgI_6$/CBTS/Au | SCAPS-1D | 1.07 | 19.62 | 82.65 | 17.47 |
| | wxAMPS | 1.18 | 21.09 | 84.73 | 21.03 |
| ITO/IGZO/$Cs_2BiAgI_6$/CBTS/Au | SCAPS-1D | 1.09 | 23.65 | 79.04 | 20.42 |
| | wxAMPS | 1.19 | 22.87 | 82.02 | 22.38 |
| ITO/$SnO_2$/$Cs_2BiAgI_6$/CBTS/Au | SCAPS-1D | 1.08 | 23.76 | 83.54 | 21.52 |
| | wxAMPS | 1.18 | 21.94 | 82.62 | 21.93 |
| ITO/$CeO_2$/$Cs_2BiAgI_6$/CBTS/Au | SCAPS-1D | 0.92 | 23.59 | 66.21 | 14.44 |
| | wxAMPS | - | - | - | - |
| ITO/$WS_2$/$Cs_2BiAgI_6$/CBTS/Au | SCAPS-1D | 1.08 | 23.53 | 83.91 | 21.43 |
| | wxAMPS | 1.18 | 21.16 | 84.94 | 21.23 |

### *3.3.2 Comparison of SCAPS-1D results with previous work*

**Table 6** compares the performance parameters of the presented eight device configurations with recently published optimum configurations. **Table 6** shows that the presented optimum $Cs_2BiAgI_6$ double perovskite-based solar cell shows a higher PCE value than the previously published $Cs_2BiAgX_6$ device structure. Presented eight sets of device structures PCE are 19.99%, 21.55%, 21.59%, 20.42%, 21.52%, 21.43%, 17.47%, and 14.44% whereas previously published device structures such as FTO/PCBM/$Cs_2BiAgI_6$/PTAA/Au structure shows almost 16.23% PCE [72], FTO/$TiO_2$/$Cs_2BiAgI_6$/Spiro-OMeTAD/Au structure shows around 2.43% PCE [91]. The $V_{OC}$ values of the presented solar structures are compatible with published device configurations. In contrast, the $J_{SC}$ and *FF* values of the presented solar structure are higher than those of the previously published $Cs_2BiAgX_6$-base device structure. All presented solar structure shows $J_{SC}$ of greater than 19 mA/cm$^{2,}$ whereas earlier published device structure shows the lowest $J_{SC}$ value except FTO/PCBM/$Cs_2BiAgI_6$/PTAA/Au structure. **Table 6** shows that the presented eight solar structures are shown more effective performance than the previously reported $Cs_2BiAgX_6$-based solar cells.

**Table 6**. The comparison of PV parameters of $Cs_2BiAgI_6$ and similar absorbers-based solar cells.

| Type | Device structure | Voc (V) | Jsc (mA/cm$^2$) | FF (%) | PCE (%) | Year | Ref. |
|---|---|---|---|---|---|---|---|
| E | FTO/c-$TiO_2$/m$TiO_2$/$Cs_2AgBiBr_6$/ N719/SpiroOMeTAD/Ag | 1.06 | 5.13 | - | 2.84 | 2020 | [92] |
| E | FTO/$TiO_2$/$Cs_2AgBiBr_6$/Spiro-OMeTAD/Au | 0.98 | 3.96 | 62.40 | 2.43 | 2021 | [91] |
| T | FTO/PCBM/$Cs_2BiAgI_6$/PTAA/Au | 1.08 | 19.94 | 74.87 | 16.23 | 2021 | [72] |
| T | ohmic contact/Spiro-OMeTAD/$Cs_2BiAgI_6$/ $TiO_2$/$SnO_2$:F($ZnO_2$)/ohmic contact | 1.18 | 16.2 | 80.20 | 15.90 | 2018 | [93] |
| T | Glass/FTO/$TiO_2$/$Cs_2AgBiBr_6$/$Cu_2O$/Au | 1.5 | 11.45 | 42.10 | 7.25 | 2020 | [90] |
| T | ZnO-NR/$Cs_2AgBiBr_6$/P3HT(Base) | 0.91 | 11.10 | 44.02 | 4.48 | 2021 | [40] |
| T | ITO/PCBM/$Cs_2BiAgI_6$/CBTS/Au | 1.08 | 22.39 | 82.56 | 19.99 | - | * |
| T | ITO/$TiO_2$/$Cs_2BiAgI_6$/CBTS/Au | 1.08 | 23.8 | 83.61 | 21.55 | - | * |
| T | ITO/ZnO/$Cs_2BiAgI_6$/CBTS/Au | 1.08 | 23.76 | 83.78 | 21.59 | - | * |
| T | ITO/$C_{60}$/$Cs_2BiAgI_6$/CBTS/Au | 1.07 | 19.62 | 82.65 | 17.47 | - | * |
| T | ITO/IGZO/$Cs_2BiAgI_6$/CBTS/Au | 1.09 | 23.65 | 79.04 | 20.42 | - | * |
| T | ITO/$SnO_2$/$Cs_2BiAgI_6$/CBTS/Au | 1.08 | 23.76 | 83.54 | 21.52 | - | * |
| T | ITO/$CeO_2$/$Cs_2BiAgI_6$/CBTS/Au | 0.92 | 23.59 | 66.21 | 14.44 | - | * |
| T | ITO/$WS_2$/$Cs_2BiAgI_6$/CBTS/Au | 1.08 | 23.53 | 83.91 | 21.43 | - | * |

Note: E = Experimental, T = Theoretical, *This work

The first 2 solar cell device structures of **Table 6** used a different absorber, i.e., $Cs_2AgBiBr_6$ (which shows less than 3% efficiency) than our studied absorber, i.e., $Cs_2BiAgI_6$. Up till now, the research on photovoltaics using a $Cs_2BiAgI_6$ absorber has been unique which was different from the experimental studied $Cs_2BiAgBr_6$ absorber-based solar cells. Our studied absorbers' characteristics like thickness, bandgap, acceptor, defect density, etc. varied from the previous theoretical and experimental studies of device structures. At the same time, our studied ETLs, and HTLs are not matched with previously studied experimental ones as per their properties as well. Furthermore, optical properties also vary from absorber to absorber which leads to solar energy absorption. The studied $Cs_2BiAgI_6$ absorber had better optical properties which were evident from the performance of 16.23% of PCE in the case of FTO/PCBM/$Cs_2BiAgI_6$/PTAA/Au structure reported by Srivastava *et al.* [72]. From all the above reasons, we can conclude that our studied $Cs_2AgBiBr_6$ solar cell shows better PCE than other studied similar solar absorbers, i.e., $Cs_2AgBiBr_6$ based different structured solar cells.

## 4 Summary

This work concludes a combined DFT, SCAPS-1D, and wxAMPS based study to design, investigate and optimize $Cs_2BiAgI_6$ double perovskite-based solar cells. Both the band structure and density of states via DFT analyses confirm the semiconducting nature of $Cs_2BiAgI_6$ material. The bonding nature of Cs-I and Ag-I bonds is found to be ionic, while it is a covalent Bi-I bond. Moreover, the analysis of optical properties suggests the possible application of this double perovskite in solar cells. In the next step, $Cs_2BiAgI_6$-based double perovskite solar cell configurations are investigated with ninety-six combinations of solar cell structures using eight ETLs and twelve HTLs through SCAPS-1D simulation. The $Cs_2BiAgI_6$ absorber layer-based device with CBTS as HTL delivered 19.99%, 21.55%, 21.59%, 17.47%, 20.42%, 21.52%, 14.44%, and 21.43% PCE with PCBM, $TiO_2$, ZnO, $C_{60}$, IGZO, $SnO_2$, $CeO_2$, $WS_2$ ETL respectively. After a comparison of various characteristics among these eight best-optimized configurations, the following findings are summarized:

1. Among all studied combinations, $TiO_2$, ZnO, and $SnO_2$ ETLs-associated devices showed comparatively better PCE of ~21.5% with CBTS HTL due to its suitable band alignment.
2. From the effect of absorber and ETL thickness, it is evident that $SnO_2$ and $WS_2$ ETLs-associated devices can deliver ~21.5% of PCE.
3. ITO/ZnO/$Cs_2BiAgI_6$/CBTS/Au PSC showed a considerable decline of PCE while series resistance increased to 6 Ω-cm$^2$.
4. In the case of shunt resistance, the performance of ITO/ZnO/$Cs_2BiAgI_6$/CBTS/Au structure is better in comparison with other devices.
5. While the temperature increased from 275 K to 475 K the performance of the ZnO ETL-associated structure is the best among all devices.
6. $V_{OC}$, and $J_{SC}$ were almost independent with respect to $R_s$, and $J_{SC}$ also remained unaffected with temperature.
7. For capacitance and Mott-Schottky characteristics, $CeO_2$ and $C_{60}$ ETLs-associated devices showed relatively maximum PCE at 0.8 V for a voltage sweep from -0.4 V to 0.8 V.
8. Also, $CeO_2$ and $C_{60}$ ETLs associated devices showed the highest generation and recombination rates at 0.9 and 1 μm respectively.
9. ZnO ETL device showed better *J-V* and *QE* characteristics due to its suitable band alignment in comparison with other ETLs associated devices.
10. Comparing SCAPS-1D results with wxAMPS results, it is evident that the difference between all the devices is marginal, which might help to understand the accuracy of the work.

Extensive numerical simulations reported in this work can be extended to other perovskite materials to study the potential perovskite absorber materials for solar cell applications, followed by development using device simulators. This can help the experimentalists to synthesize the desired material with the most efficient device architecture.

**Author's contributions**

**M.K. Hossain**: Conceptualization, Methodology, Software, Validation, Formal analysis, Investigation, Data curation, Writing – original draft, Writing – review & editing, Supervision, Project administration; **A.A. Arnab**: Formal analysis, Investigation, Data curation, Writing – original draft; **R.C. Das**: Writing – review & editing;

**K.M. Hossain**: Writing – original draft; **M.H.K. Rubel**: Software, Data curation; Formal analysis, Investigation, Writing – review & editing; **M.F. Rahman**: Writing – review & editing; **H. Bencherif**: Writing – review & editing; **M. E. Emetere**: Writing – review & editing; **M.K.A. Mohammed**: Writing – review & editing; **R. Pandey**: Writing – review & editing.

**Data availability**

The raw/processed data required to reproduce these findings cannot be shared at this time as the data also forms part of an ongoing study.

**Declaration of interests**

The authors declare that they have no known competing financial interests or personal relationships that could have appeared to influence the work reported in this paper.

**Funding**

This research did not receive any specific grant from funding agencies in the public, commercial, or not-for-profit sectors.

**Acknowledgments**

The SCAPS-1D program was provided by Dr. M. Burgelman of the University of Gent in Belgium. The authors would like to express their gratitude to him. They would also like to thank Professor A. Rockett and Dr. Yiming Liu from UIUC, as well as Professor Fonash from Penn State University, for their contributions to the wxAMPS program. We also acknowledge G.F. Ishraque Toki of Donghua University, Shanghai, China for reviewing the manuscript.

**References**

[1] H. Tang, S. He, C. Peng, A Short Progress Report on High-Efficiency Perovskite Solar Cells, Nanoscale Res. Lett. 12 (2017) 410. https://doi.org/10.1186/s11671-017-2187-5.

[2] M.K. Nazeeruddin, Twenty-five years of low-cost solar cells, Nature. 538 (2016) 463–464. https://doi.org/10.1038/538463a.

[3] M.F. Rahman, M.J.A. Habib, M.H. Ali, M.H.K. Rubel, M.R. Islam, A.B. Md. Ismail, M.K. Hossain, Design and numerical investigation of cadmium telluride (CdTe) and iron silicide (FeSi 2 ) based double absorber solar cells to enhance power conversion efficiency, AIP Adv. 12 (2022) 105317. https://doi.org/10.1063/5.0108459.

[4] M.M.M. Islam, A. Kowsar, A.K.M.M. Haque, M.K. Hossain, M.H. Ali, M.H.K. Rubel, M.F. Rahman, Techno-economic Analysis of Hybrid Renewable Energy System for Healthcare Centre in Northwest Bangladesh, Process Integr. Optim. Sustain. (2022). https://doi.org/10.1007/s41660-022-00294-8.

[5] M.K. Hossain, S.M.K. Hasan, M.I. Hossain, R.C. Das, H. Bencherif, M.H.K. Rubel, M.F. Rahman, T. Emrose, K. Hashizume, A Review of Applications, Prospects, and Challenges of Proton-Conducting Zirconates in Electrochemical Hydrogen Devices, Nanomaterials. 12 (2022) 3581. https://doi.org/10.3390/nano12203581.

[6] M.K. Hossain, G.A. Raihan, M.A. Akbar, M.H. Kabir Rubel, M.H. Ahmed, M.I. Khan, S. Hossain, S.K. Sen, M.I.E. Jalal, A. El-Denglawey, Current Applications and Future Potential of Rare Earth Oxides in Sustainable Nuclear, Radiation, and Energy Devices: A Review, ACS Appl. Electron. Mater. 4 (2022) 3327–3353. https://doi.org/10.1021/acsaelm.2c00069.

[7] M.K. Hossain, M.H.K. Rubel, M.A. Akbar, M.H. Ahmed, N. Haque, M.F. Rahman, J. Hossain, K.M. Hossain, A review on recent applications and future prospects of rare earth oxides in corrosion and thermal barrier coatings, catalysts, tribological, and environmental sectors, Ceram. Int. 48 (2022) 32588–32612. https://doi.org/10.1016/j.ceramint.2022.07.220.

[8] M.K. Hossain, R. Chanda, A. El-Denglawey, T. Emrose, M.T. Rahman, M.C. Biswas, K. Hashizume, Recent progress in barium zirconate proton conductors for electrochemical hydrogen device applications: A review, Ceram. Int. 47 (2021) 23725–23748. https://doi.org/10.1016/j.ceramint.2021.05.167.

[9] A. Kojima, K. Teshima, Y. Shirai, T. Miyasaka, Organometal Halide Perovskites as Visible-Light Sensitizers for Photovoltaic Cells, J. Am. Chem. Soc. 131 (2009) 6050–6051. https://doi.org/10.1021/ja809598r.

[10] Y. Chen, L. Zhang, Y. Zhang, H. Gao, H. Yan, Large-area perovskite solar cells – a review of recent

progress and issues, RSC Adv. 8 (2018) 10489–10508. https://doi.org/10.1039/C8RA00384J.

[11] C.O. Ramírez Quiroz, Y. Shen, M. Salvador, K. Forberich, N. Schrenker, G.D. Spyropoulos, T. Heumüller, B. Wilkinson, T. Kirchartz, E. Spiecker, P.J. Verlinden, X. Zhang, M.A. Green, A. Ho-Baillie, C.J. Brabec, Balancing electrical and optical losses for efficient 4-terminal Si–perovskite solar cells with solution processed percolation electrodes, J. Mater. Chem. A. 6 (2018) 3583–3592. https://doi.org/10.1039/C7TA10945H.

[12] K. Yoshikawa, H. Kawasaki, W. Yoshida, T. Irie, K. Konishi, K. Nakano, T. Uto, D. Adachi, M. Kanematsu, H. Uzu, K. Yamamoto, Silicon heterojunction solar cell with interdigitated back contacts for a photoconversion efficiency over 26%, Nat. Energy. 2 (2017) 17032. https://doi.org/10.1038/nenergy.2017.32.

[13] N. Kumar, J. Rani, R. Kurchania, Advancement in CsPbBr3 inorganic perovskite solar cells: Fabrication, efficiency and stability, Sol. Energy. 221 (2021) 197–205. https://doi.org/10.1016/j.solener.2021.04.042.

[14] M.K. Hossain, M.H.K. Rubel, G.F.I. Toki, I. Alam, M.F. Rahman, H. Bencherif, Effect of Various Electron and Hole Transport Layers on the Performance of CsPbI 3 -Based Perovskite Solar Cells: A Numerical Investigation in DFT, SCAPS-1D, and wxAMPS Frameworks, ACS Omega. 7 (2022) 43210–43230. https://doi.org/10.1021/acsomega.2c05912.

[15] Y. Raoui, H. Ez-Zahraouy, S. Ahmad, S. Kazim, Unravelling the theoretical window to fabricate high performance inorganic perovskite solar cells, Sustain. Energy Fuels. 5 (2021) 219–229. https://doi.org/10.1039/D0SE01160F.

[16] A. Babayigit, A. Ethirajan, M. Muller, B. Conings, Toxicity of organometal halide perovskite solar cells, Nat. Mater. 15 (2016) 247–251. https://doi.org/10.1038/nmat4572.

[17] I.R. Benmessaoud, A.-L. Mahul-Mellier, E. Horváth, B. Maco, M. Spina, H.A. Lashuel, L. Forró, Health hazards of methylammonium lead iodide based perovskites: cytotoxicity studies, Toxicol. Res. (Camb). 5 (2016) 407–419. https://doi.org/10.1039/C5TX00303B.

[18] A. Babayigit, D. Duy Thanh, A. Ethirajan, J. Manca, M. Muller, H.-G. Boyen, B. Conings, Assessing the toxicity of Pb- and Sn-based perovskite solar cells in model organism Danio rerio, Sci. Rep. 6 (2016) 18721. https://doi.org/10.1038/srep18721.

[19] H. Bencherif, M.K. Hossain, Design and numerical investigation of efficient (FAPbI3)1−x(CsSnI3)x perovskite solar cell with optimized performances, Sol. Energy. 248 (2022) 137–148. https://doi.org/10.1016/j.solener.2022.11.012.

[20] H. Bencherif, F. Meddour, M.H. Elshorbagy, M.K. Hossain, A. Cuadrado, M.A. Abdi, T. Bendib, S. Kouda, J. Alda, Performance enhancement of (FAPbI3)1-x(MAPbBr3)x perovskite solar cell with an optimized design, Micro and Nanostructures. 171 (2022) 207403. https://doi.org/10.1016/j.micrna.2022.207403.

[21] O.A. Lozhkina, A.A. Murashkina, M.S. Elizarov, V.V. Shilovskikh, A.A. Zolotarev, Y. V. Kapitonov, R. Kevorkyants, A.V. Emeline, T. Miyasaka, Microstructural analysis and optical properties of the halide double perovskite Cs2BiAgBr6 single crystals, Chem. Phys. Lett. 694 (2018) 18–22. https://doi.org/10.1016/j.cplett.2018.01.031.

[22] R.E. Brandt, R.C. Kurchin, R.L.Z. Hoye, J.R. Poindexter, M.W.B. Wilson, S. Sulekar, F. Lenahan, P.X.T. Yen, V. Stevanović, J.C. Nino, M.G. Bawendi, T. Buonassisi, Investigation of Bismuth Triiodide (BiI 3 ) for Photovoltaic Applications, J. Phys. Chem. Lett. 6 (2015) 4297–4302. https://doi.org/10.1021/acs.jpclett.5b02022.

[23] H. Zhou, Q. Chen, G. Li, S. Luo, T. Song, H.-S. Duan, Z. Hong, J. You, Y. Liu, Y. Yang, Interface engineering of highly efficient perovskite solar cells, Science (80-. ). 345 (2014) 542–546. https://doi.org/10.1126/science.1254050.

[24] F. Giustino, H.J. Snaith, Toward Lead-Free Perovskite Solar Cells, ACS Energy Lett. 1 (2016) 1233–1240. https://doi.org/10.1021/acsenergylett.6b00499.

[25] F. Jiang, D. Yang, Y. Jiang, T. Liu, X. Zhao, Y. Ming, B. Luo, F. Qin, J. Fan, H. Han, L. Zhang, Y. Zhou, Chlorine-Incorporation-Induced Formation of the Layered Phase for Antimony-Based Lead-Free Perovskite Solar Cells, J. Am. Chem. Soc. 140 (2018) 1019–1027. https://doi.org/10.1021/jacs.7b10739.

[26] E.T. McClure, M.R. Ball, W. Windl, P.M. Woodward, Cs 2 AgBiX 6 (X = Br, Cl): New Visible Light Absorbing, Lead-Free Halide Perovskite Semiconductors, Chem. Mater. 28 (2016) 1348–1354. https://doi.org/10.1021/acs.chemmater.5b04231.

[27] P.-K. Kung, M.-H. Li, P.-Y. Lin, J.-Y. Jhang, M. Pantaler, D.C. Lupascu, G. Grancini, P. Chen, Lead-Free Double Perovskites for Perovskite Solar Cells, Sol. RRL. 4 (2020) 1900306. https://doi.org/10.1002/solr.201900306.

[28] K. Dave, M.H. Fang, Z. Bao, H.T. Fu, R.S. Liu, Recent Developments in Lead-Free Double Perovskites: Structure, Doping, and Applications, Chem. - An Asian J. 15 (2020) 242–252. https://doi.org/10.1002/asia.201901510.

[29] C. Frangville, M. Rutkevičius, A.P. Richter, O.D. Velev, S.D. Stoyanov, V.N. Paunov, Fabrication of Environmentally Biodegradable Lignin Nanoparticles, ChemPhysChem. 13 (2012) 4235–4243. https://doi.org/10.1002/cphc.201200537.

[30] P.K. Nayak, D.T. Moore, B. Wenger, S. Nayak, A.A. Haghighirad, A. Fineberg, N.K. Noel, O.G. Reid, G. Rumbles, P. Kukura, K.A. Vincent, H.J. Snaith, Mechanism for rapid growth of organic–inorganic halide perovskite crystals, Nat. Commun. 7 (2016) 13303. https://doi.org/10.1038/ncomms13303.

[31] F. Marchetti, E. Moroni, A. Pandini, G. Colombo, Machine Learning Prediction of Allosteric Drug Activity from Molecular Dynamics, J. Phys. Chem. Lett. 12 (2021) 3724–3732. https://doi.org/10.1021/acs.jpclett.1c00045.

[32] A.H. Slavney, T. Hu, A.M. Lindenberg, H.I. Karunadasa, A Bismuth-Halide Double Perovskite with Long Carrier Recombination Lifetime for Photovoltaic Applications, J. Am. Chem. Soc. 138 (2016) 2138–2141. https://doi.org/10.1021/jacs.5b13294.

[33] W. Ning, F. Wang, B. Wu, J. Lu, Z. Yan, X. Liu, Y. Tao, J.-M. Liu, W. Huang, M. Fahlman, L. Hultman, T.C. Sum, F. Gao, Long Electron-Hole Diffusion Length in High-Quality Lead-Free Double Perovskite Films, Adv. Mater. 30 (2018) 1706246. https://doi.org/10.1002/adma.201706246.

[34] G. Volonakis, M.R. Filip, A.A. Haghighirad, N. Sakai, B. Wenger, H.J. Snaith, F. Giustino, Lead-Free Halide Double Perovskites via Heterovalent Substitution of Noble Metals, J. Phys. Chem. Lett. 7 (2016) 1254–1259. https://doi.org/10.1021/acs.jpclett.6b00376.

[35] F. Igbari, Z.K. Wang, L.S. Liao, Progress of Lead-Free Halide Double Perovskites, Adv. Energy Mater. 9 (2019) 1–32. https://doi.org/10.1002/aenm.201803150.

[36] X.-G. Zhao, D. Yang, J.-C. Ren, Y. Sun, Z. Xiao, L. Zhang, Rational Design of Halide Double Perovskites for Optoelectronic Applications, Joule. 2 (2018) 1662–1673. https://doi.org/10.1016/j.joule.2018.06.017.

[37] H. Absike, N. Baaalla, R. Lamouri, H. Labrim, H. Ez-zahraouy, Optoelectronic and photovoltaic properties of Cs2AgBiX6 (X = Br, Cl, or I) halide double perovskite for solar cells: Insight from density functional theory, Int. J. Energy Res. 46 (2022) 11053–11064. https://doi.org/10.1002/er.7907.

[38] H. Wu, A. Erbing, M.B. Johansson, J. Wang, C. Kamal, M. Odelius, E.M.J. Johansson, Mixed-Halide Double Perovskite Cs 2 AgBiX 6 (X=Br, I) with Tunable Optical Properties via Anion Exchange, ChemSusChem. 14 (2021) 4507–4515. https://doi.org/10.1002/cssc.202101146.

[39] N. Singh, A. Agarwal, M. Agarwal, Performance evaluation of lead–free double-perovskite solar cell, Opt. Mater. (Amst). 114 (2021) 110964. https://doi.org/10.1016/j.optmat.2021.110964.

[40] I. Alam, R. Mollick, M.A. Ashraf, Numerical simulation of Cs2AgBiBr6-based perovskite solar cell with ZnO nanorod and P3HT as the charge transport layers, Phys. B Condens. Matter. 618 (2021) 413187. https://doi.org/10.1016/j.physb.2021.413187.

[41] J. Gong, S.B. Darling, F. You, Perovskite photovoltaics: life-cycle assessment of energy and environmental impacts, Energy Environ. Sci. 8 (2015) 1953–1968. https://doi.org/10.1039/C5EE00615E.

[42] H. Choi, S. Park, S. Paek, P. Ekanayake, M.K. Nazeeruddin, J. Ko, Efficient star-shaped hole transporting materials with diphenylethenyl side arms for an efficient perovskite solar cell, J. Mater. Chem. A. 2 (2014) 19136–19140. https://doi.org/10.1039/C4TA04179H.

[43] C. Devi, R. Mehra, Device simulation of lead-free MASnI3 solar cell with CuSbS2 (copper antimony sulfide), J. Mater. Sci. 54 (2019) 5615–5624. https://doi.org/10.1007/s10853-018-03265-y.

[44] M. Liu, M.B. Johnston, H.J. Snaith, Efficient planar heterojunction perovskite solar cells by vapour deposition, Nature. 501 (2013) 395–398. https://doi.org/10.1038/nature12509.

[45] M.K. Hossain, M.T. Rahman, M.K. Basher, M.S. Manir, M.S. Bashar, Influence of thickness variation of gamma-irradiated DSSC photoanodic TiO2 film on structural, morphological and optical properties, Optik (Stuttg). 178 (2019) 449–460. https://doi.org/10.1016/j.ijleo.2018.09.170.

[46] M.K. Hossain, M.T. Rahman, M.K. Basher, M.J. Afzal, M.S. Bashar, Impact of ionizing radiation doses on nanocrystalline TiO2 layer in DSSC's photoanode film, Results Phys. 11 (2018) 1172–1181. https://doi.org/10.1016/j.rinp.2018.10.006.

[47] M.K. Hossain, A.A. Mortuza, S.K. Sen, M.K. Basher, M.W. Ashraf, S. Tayyaba, M.N.H. Mia, M.J. Uddin, A comparative study on the influence of pure anatase and Degussa-P25 TiO2 nanomaterials on the structural and optical properties of dye sensitized solar cell (DSSC) photoanode, Optik (Stuttg). 171 (2018) 507–516. https://doi.org/10.1016/j.ijleo.2018.05.032.

[48] M.K. Hossain, M.F. Pervez, M.N.H. Mia, A.A. Mortuza, M.S. Rahaman, M.R. Karim, J.M.M. Islam, F. Ahmed, M.A. Khan, Effect of dye extracting solvents and sensitization time on photovoltaic performance of natural dye sensitized solar cells, Results Phys. 7 (2017) 1516–1523. https://doi.org/10.1016/j.rinp.2017.04.011.

[49] M.K. Hossain, M.F. Pervez, M.J. Uddin, S. Tayyaba, M.N.H. Mia, M.S. Bashar, M.K.H. Jewel, M.A.S. Haque, M.A. Hakim, M.A. Khan, Influence of natural dye adsorption on the structural, morphological and optical properties of TiO 2 based photoanode of dye-sensitized solar cell, Mater. Sci. 36 (2017) 93–101. https://doi.org/10.1515/msp-2017-0090.

[50] M.K. Hossain, M.F. Pervez, M.N.H. Mia, S. Tayyaba, M.J. Uddin, R. Ahamed, R.A. Khan, M. Hoq, M.A. Khan, F. Ahmed, Annealing temperature effect on structural, morphological and optical parameters of mesoporous TiO 2 film photoanode for dye-sensitized solar cell application, Mater. Sci. 35 (2017) 868–877. https://doi.org/10.1515/msp-2017-0082.

[51] M.K. Hossain, M.F. Pervez, S. Tayyaba, M.J. Uddin, A.A. Mortuza, M.N.H. Mia, M.S. Manir, M.R. Karim, M.A. Khan, Efficiency enhancement of natural dye sensitized solar cell by optimizing electrode fabrication parameters, Mater. Sci. 35 (2017) 816–823. https://doi.org/10.1515/msp-2017-0086.

[52] Y. Raoui, H. Ez-Zahraouy, N. Tahiri, O. El Bounagui, S. Ahmad, S. Kazim, Performance analysis of MAPbI3 based perovskite solar cells employing diverse charge selective contacts: Simulation study, Sol. Energy. 193 (2019) 948–955. https://doi.org/10.1016/j.solener.2019.10.009.

[53] S.Z. Haider, H. Anwar, M. Wang, A comprehensive device modelling of perovskite solar cell with inorganic copper iodide as hole transport material, Semicond. Sci. Technol. 33 (2018) 035001. https://doi.org/10.1088/1361-6641/aaa596.

[54] M.H.K. Rubel, M.A. Hossain, M.K. Hossain, K.M. Hossain, A.A. Khatun, M.M. Rahaman, M. Ferdous Rahman, M.M. Hossain, J. Hossain, First-principles calculations to investigate structural, elastic, electronic, thermodynamic, and thermoelectric properties of CaPd3B4O12 (B = Ti, V) perovskites, Results Phys. 42 (2022) 105977. https://doi.org/10.1016/j.rinp.2022.105977.

[55] M.H.K. Rubel, S.K. Mitro, M.K. Hossain, K.M. Hossain, M.M. Rahaman, J. Hossain, B.K. Mondal, A. Akter, M.F. Rahman, I. Ahmed, A.K.M.A. Islam, First-principles calculations to investigate physical properties of single-cubic (Ba0.82K0.18)(Bi0.53Pb0.47)O3 novel perovskite superconductor, Mater. Today Commun. 33 (2022) 104302. https://doi.org/10.1016/j.mtcomm.2022.104302.

[56] M.I. Kholil, M.T.H. Bhuiyan, M.A. Rahman, M.S. Ali, M. Aftabuzzaman, Effects of Fe doping on the visible light absorption and bandgap tuning of lead-free (CsSnCl 3 ) and lead halide (CsPbCl 3 ) perovskites for optoelectronic applications, AIP Adv. 11 (2021) 035229. https://doi.org/10.1063/5.0042847.

[57] M. Roknuzzaman, K. Ostrikov, H. Wang, A. Du, T. Tesfamichael, Towards lead-free perovskite photovoltaics and optoelectronics by ab-initio simulations, Sci. Rep. 7 (2017) 14025. https://doi.org/10.1038/s41598-017-13172-y.

[58] J. Ur Rehman, M. Usman, S. Amjid, M. Sagir, M. Bilal Tahir, A. Hussain, I. Alam, R. Nazir, H. Alrobei, S. Ullah, M. Ali Assiri, First-principles calculations to investigate structural, electronics, optical and elastic properties of Sn-based inorganic Halide-perovskites CsSnX3 (X = I, Br, Cl) for solar cell applications, Comput. Theor. Chem. 1209 (2022) 113624. https://doi.org/10.1016/j.comptc.2022.113624.

[59] M.A. Hadi, M.N. Islam, J. Podder, Indirect to direct band gap transition through order to disorder transformation of Cs 2 AgBiBr 6 via creating antisite defects for optoelectronic and photovoltaic applications, RSC Adv. 12 (2022) 15461–15469. https://doi.org/10.1039/D1RA06308A.

[60] J. Zhou, X. Rong, M.S. Molokeev, X. Zhang, Z. Xia, Exploring the transposition effects on the electronic and optical properties of Cs 2 AgSbCl 6 via a combined computational-experimental approach, J. Mater. Chem. A. 6 (2018) 2346–2352. https://doi.org/10.1039/C7TA10062K.

[61] M.D. Segall, P.J.D. Lindan, M.J. Probert, C.J. Pickard, P.J. Hasnip, S.J. Clark, M.C. Payne, First-principles simulation: ideas, illustrations and the CASTEP code, J. Phys. Condens. Matter. 14 (2002) 2717–2744. https://doi.org/10.1088/0953-8984/14/11/301.

[62] M.C. Payne, M.P. Teter, D.C. Allan, T.A. Arias, J.D. Joannopoulos, Iterative minimization techniques for ab initio total-energy calculations: molecular dynamics and conjugate gradients, Rev. Mod. Phys. 64 (1992) 1045–1097. https://doi.org/10.1103/RevModPhys.64.1045.

[63] D. Vanderbilt, Soft self-consistent pseudopotentials in a generalized eigenvalue formalism, Phys. Rev. B. 41 (1990) 7892–7895. https://doi.org/10.1103/PhysRevB.41.7892.

[64] J.P. Perdew, K. Burke, M. Ernzerhof, Generalized Gradient Approximation Made Simple, Phys. Rev. Lett. 77 (1996) 3865–3868. https://doi.org/10.1103/PhysRevLett.77.3865.

[65] H.J. Monkhorst, J.D. Pack, Special points for Brillouin-zone integrations, Phys. Rev. B. 13 (1976) 5188–5192. https://doi.org/10.1103/PhysRevB.13.5188.

[66] T.H. Fischer, J. Almlof, General methods for geometry and wave function optimization, J. Phys. Chem. 96 (1992) 9768–9774. https://doi.org/10.1021/j100203a036.

[67] Z. Boekelheide, T. Saerbeck, A.P.J. Stampfl, R.A. Robinson, D.A. Stewart, F. Hellman, Antiferromagnetism in Cr 3 Al and relation to semiconducting behavior, Phys. Rev. B. 85 (2012) 094413. https://doi.org/10.1103/PhysRevB.85.094413.

[68] S.S. Hussain, S. Riaz, G.A. Nowsherwan, K. Jahangir, A. Raza, M.J. Iqbal, I. Sadiq, S.M. Hussain, S. Naseem, Numerical Modeling and Optimization of Lead-Free Hybrid Double Perovskite Solar Cell by Using SCAPS-1D, J. Renew. Energy. 2021 (2021) 1–12. https://doi.org/10.1155/2021/6668687.

[69] F. Liu, J. Zhu, J. Wei, Y. Li, M. Lv, S. Yang, B. Zhang, J. Yao, S. Dai, Numerical simulation: Toward the design of high-efficiency planar perovskite solar cells, Appl. Phys. Lett. 104 (2014) 253508. https://doi.org/10.1063/1.4885367.

[70] Y. Liu, Y. Sun, A. Rockett, A new simulation software of solar cells - WxAMPS, Sol. Energy Mater. Sol. Cells. 98 (2012) 124–128. https://doi.org/10.1016/j.solmat.2011.10.010.

[71] J. He, T. Zhang, P. Tang, C. Qiu, X. Feng, X. Luo, Simulation on the carrier transport process inside the semiconductor of dye sensitized solar cells by wxAMPS software, Electrochim. Acta. 125 (2014) 218–224. https://doi.org/10.1016/j.electacta.2014.01.019.

[72] S. Srivastava, A.K. Singh, P. Kumar, B. Pradhan, Comparative performance analysis of lead-free perovskites solar cells by numerical simulation, J. Appl. Phys. 131 (2022) 175001. https://doi.org/10.1063/5.0088099.

[73] M.H.K. Rubel, K.M. Hossain, S.K. Mitro, M.M. Rahaman, M.A. Hadi, A.K.M.A. Islam, Comprehensive first-principles calculations on physical properties of ScV2Ga4 and ZrV2Ga4 in comparison with superconducting HfV2Ga4, Mater. Today Commun. 24 (2020) 100935. https://doi.org/10.1016/j.mtcomm.2020.100935.

[74] W.-C. Hu, Y. Liu, D.-J. Li, X.-Q. Zeng, C.-S. Xu, First-principles study of structural and electronic properties of C14-type Laves phase Al2Zr and Al2Hf, Comput. Mater. Sci. 83 (2014) 27–34. https://doi.org/10.1016/j.commatsci.2013.10.029.

[75] J. Zhou, Z. Xia, M.S. Molokeev, X. Zhang, D. Peng, Q. Liu, Composition design, optical gap and stability investigations of lead-free halide double perovskite Cs 2 AgInCl 6, J. Mater. Chem. A. 5 (2017) 15031–15037. https://doi.org/10.1039/C7TA04690A.

[76] A.H. Reshak, V. V Atuchin, S. Auluck, I. V Kityk, First and second harmonic generation of the optical susceptibilities for the non-centro-symmetric orthorhombic AgCd 2 GaS 4, J. Phys. Condens. Matter. 20 (2008) 325234. https://doi.org/10.1088/0953-8984/20/32/325234.

[77] T. Kim, J. Lim, S. Song, Recent Progress and Challenges of Electron Transport Layers in Organic–Inorganic Perovskite Solar Cells, Energies. 13 (2020) 5572. https://doi.org/10.3390/en13215572.

[78] Y. Raoui, H. Ez-Zahraouy, S. Kazim, S. Ahmad, Energy level engineering of charge selective contact and halide perovskite by modulating band offset: Mechanistic insights, J. Energy Chem. 54 (2021) 822–829. https://doi.org/10.1016/j.jechem.2020.06.030.

[79] S.B. Selssabil, Zerarka, Study of electron transport effect on perovskite solar cells using simulation, University Mohamed Khider de Biskra, 2019. http://archives.univ-biskra.dz/bitstream/123456789/15523/1/Zerarka_Selssabil_et_Seridji_Baya.pdf.

[80] A. Sunny, S. Rahman, M.M. Khatun, S.R. Al Ahmed, Numerical study of high performance HTL-free CH 3 NH 3 SnI 3 -based perovskite solar cell by SCAPS-1D, AIP Adv. 11 (2021) 065102. https://doi.org/10.1063/5.0049646.

[81] S. Karthick, S. Velumani, J. Bouclé, Experimental and SCAPS simulated formamidinium perovskite solar cells: A comparison of device performance, Sol. Energy. 205 (2020) 349–357. https://doi.org/10.1016/j.solener.2020.05.041.

[82] K. Tvingstedt, L. Gil-Escrig, C. Momblona, P. Rieder, D. Kiermasch, M. Sessolo, A. Baumann, H.J. Bolink, V. Dyakonov, Removing Leakage and Surface Recombination in Planar Perovskite Solar Cells, ACS Energy Lett. 2 (2017) 424–430. https://doi.org/10.1021/acsenergylett.6b00719.

[83] Y. Li, B. Ding, Q.-Q. Chu, G.-J. Yang, M. Wang, C.-X. Li, C.-J. Li, Ultra-high open-circuit voltage of perovskite solar cells induced by nucleation thermodynamics on rough substrates, Sci. Rep. 7 (2017) 46141. https://doi.org/10.1038/srep46141.

[84] Y. Li, Z. Shi, L. Lei, F. Zhang, Z. Ma, D. Wu, T. Xu, Y. Tian, Y. Zhang, G. Du, C. Shan, X. Li, Highly

Stable Perovskite Photodetector Based on Vapor-Processed Micrometer-Scale CsPbBr 3 Microplatelets, Chem. Mater. 30 (2018) 6744–6755. https://doi.org/10.1021/acs.chemmater.8b02435.

[85] Z. Ma, Z. Shi, C. Qin, M. Cui, D. Yang, X. Wang, L. Wang, X. Ji, X. Chen, J. Sun, D. Wu, Y. Zhang, X.J. Li, L. Zhang, C. Shan, Stable Yellow Light-Emitting Devices Based on Ternary Copper Halides with Broadband Emissive Self-Trapped Excitons, ACS Nano. 14 (2020) 4475–4486. https://doi.org/10.1021/acsnano.9b10148.

[86] Z. Shi, S. Li, Y. Li, H. Ji, X. Li, D. Wu, T. Xu, Y. Chen, Y. Tian, Y. Zhang, C. Shan, G. Du, Strategy of Solution-Processed All-Inorganic Heterostructure for Humidity/Temperature-Stable Perovskite Quantum Dot Light-Emitting Diodes, ACS Nano. 12 (2018) 1462–1472. https://doi.org/10.1021/acsnano.7b07856.

[87] S. Lin, Computer Solutions of the Traveling Salesman Problem, Bell Syst. Tech. J. 44 (1965) 2245–2269. https://doi.org/10.1002/j.1538-7305.1965.tb04146.x.

[88] M. Fischer, K. Tvingstedt, A. Baumann, V. Dyakonov, Doping Profile in Planar Hybrid Perovskite Solar Cells Identifying Mobile Ions, ACS Appl. Energy Mater. 1 (2018) acsaem.8b01119. https://doi.org/10.1021/acsaem.8b01119.

[89] S.C. Baker-Finch, K.R. McIntosh, D. Yan, K.C. Fong, T.C. Kho, Near-infrared free carrier absorption in heavily doped silicon, J. Appl. Phys. 116 (2014) 063106. https://doi.org/10.1063/1.4893176.

[90] M.T. Islam, M.R. Jani, S.M. Al Amin, M.S.U. Sami, K.M. Shorowordi, M.I. Hossain, M. Devgun, S. Chowdhury, S. Banerje, S. Ahmed, Numerical simulation studies of a fully inorganic Cs2AgBiBr6 perovskite solar device, Opt. Mater. (Amst). 105 (2020) 109957. https://doi.org/10.1016/j.optmat.2020.109957.

[91] M.S. Shadabroo, H. Abdizadeh, M.R. Golobostanfard, Dimethyl Sulfoxide Vapor-Assisted Cs 2 AgBiBr 6 Homogenous Film Deposition for Solar Cell Application, ACS Appl. Energy Mater. 4 (2021) 6797–6805. https://doi.org/10.1021/acsaem.1c00894.

[92] X. Yang, Y. Chen, P. Liu, H. Xiang, W. Wang, R. Ran, W. Zhou, Z. Shao, Simultaneous Power Conversion Efficiency and Stability Enhancement of Cs 2 AgBiBr 6 Lead-Free Inorganic Perovskite Solar Cell through Adopting a Multifunctional Dye Interlayer, Adv. Funct. Mater. 30 (2020) 2001557. https://doi.org/10.1002/adfm.202001557.

[93] B. Smith, Efficient Lead-Free Perovskite Solar Cell, Urbana, IL 61801, USA, 2018. https://443.ece.illinois.edu/files/2018/09/SolarCellReportSmithWaterMarked.pdf.